\documentclass[12pt]{article}
\usepackage[T1]{fontenc}
\usepackage[utf8]{inputenc}
\usepackage{geometry}
\usepackage{color}
\usepackage{mathtools}
\usepackage{bm}
\usepackage{amsmath}
\usepackage{amssymb}
\usepackage{graphicx}
\usepackage{setspace}
\usepackage{wasysym}
\usepackage[authoryear]{natbib}
\PassOptionsToPackage{normalem}{ulem}
\usepackage{ulem}
\usepackage[unicode=true,pdfusetitle,
 bookmarks=true,bookmarksnumbered=false,bookmarksopen=false,
 breaklinks=false,pdfborder={0 0 1},backref=false,colorlinks=true]
 {hyperref}
\hypersetup{
 linkcolor=blue, citecolor=blue, pdfstartview={FitH}, hyperfootnotes=false, unicode=true}

\makeatletter

\usepackage{chngcntr}
\usepackage{mathtools}
\counterwithout{figure}{section}
\counterwithout{equation}{section}
\usepackage{times}
\usepackage{subfig}
\usepackage{amsmath}
\usepackage{floatrow}
\usepackage[font=small]{caption}
\date{}

\usepackage{hyperref}
\bibpunct{(}{)}{,}{a}{,}{,}

\makeatother

\global\long\def\nthc#1{h^{\left(#1\right)}}%
\global\long\def\nrho#1{\rho^{\left(#1\right)}}%
\global\long\def\nmu#1{\mu^{\left(#1\right)}}%
\global\long\def\nlambda#1{\lambda^{\left(#1\right)}}%
\global\long\def\ncL#1{c_{L}^{\left(#1\right)}}%
\global\long\def\ncLSquare#1{c_{L}^{2\left(#1\right)}}%
\global\long\def\ncS#1{c_{S}^{\left(#1\right)}}%
\global\long\def\ncSSquare#1{c_{S}^{2\left(#1\right)}}%
\global\long\def\ksnell{k_{2}}%
\global\long\def\kLSquare#1{k_{L}^{2\left(#1\right)}}%
\global\long\def\kSSquare#1{k_{S}^{2\left(#1\right)}}%
\global\long\def\ALp#1{A_{L+}^{\left(#1\right)}}%
\global\long\def\ALm#1{A_{L-}^{\left(#1\right)}}%
\global\long\def\ASp#1{A_{S+}^{\left(#1\right)}}%
\global\long\def\ASm#1{A_{S-}^{\left(#1\right)}}%
\global\long\def\kL#1{k_{L}^{\left(#1\right)}}%
\global\long\def\kS#1{k_{S}^{\left(#1\right)}}%
\global\long\def\ub{\boldsymbol{u}}%
\global\long\def\sigmab{\boldsymbol{\sigma}}%
\global\long\def\stateBC#1{\mathsf{s}_{#1}^{\perp}}%
\global\long\def\state{\mathsf{s}}%
\global\long\def\sm{\mathsf{s_{m}}}%
\global\long\def\kSAVG{\bar{k_{S}}}%
\global\long\def\kLAVG{\bar{k_{L}}}%
\global\long\def\muAVG{\bar{\mu}}%
\global\long\def\lambdaAVG{\bar{\lambda}}%
\global\long\def\Qmat{\mathsf{Q}}%
\global\long\def\nh#1{h^{\left(#1\right)}}%
\global\long\def\nH#1{\mathsf{H}^{\left(#1\right)}}%
\global\long\def\Hmat{\mathsf{H}}%
\global\long\def\kb{k_{\mathrm{B}}}%
\global\long\def\IL{I}%
\global\long\def\T{\mathbf{T}}%
\global\long\def\Tm#1{T_{#1}}%
\global\long\def\km{k_{m}}%
\global\long\def\k{k}%
\global\long\def\phiR{\phi_{R}}%
\global\long\def\psiR{\psi_{R}}%
\global\long\def\Rnphi#1{R_{#1}^{\phi}}%
\global\long\def\Rnpsi#1{R_{#1}^{\psi}}%
\global\long\def\knpsi#1{\kappa_{#1}^{\psi}}%
\global\long\def\knphi#1{\kappa_{#1}^{\phi}}%
\global\long\def\U#1{U_{#1}\left(x_{1}\right)}%
\global\long\def\NR{N_{R}}%
\global\long\def\nth{n^{\mathrm{th}}}%
\global\long\def\avgP#1{\left\langle \mathcal{P}_{#1}\right\rangle }%
\global\long\def\Cu#1{C_{u_{#1}}}%
\global\long\def\Csig#1{C_{\sigma_{#1}}}%
\global\long\def\Zeros#1{\mathsf{0}_{#1}}%
\global\long\def\Ones#1{\mathsf{I}_{#1}}%
\global\long\def\a#1{a_{#1}}%
\global\long\def\P#1{\mathcal{P}_{#1}}%
\global\long\def\bH{\mathsf{H}}%
\global\long\def\bI{\mathsf{I}_{2}}%
\global\long\def\dupm#1{\tilde{u}'_{#1}}%
\global\long\def\upm#1{\tilde{u}_{#1}}%
\global\long\def\conjupm#1{\tilde{u}_{#1}^{*}}%
\global\long\def\upmb#1{\boldsymbol{\tilde{u}}_{#1}}%
\global\long\def\prop{\zeta}%
\global\long\def\G{\mathrm{G}}%
\global\long\def\GTag{\mathrm{G'}}%
\global\long\def\GTTag{\mathrm{G''}}%
\global\long\def\Bmat{\mathsf{B}}%
\global\long\def\Amat{\mathsf{A}}%
\global\long\def\nprop#1{\prop^{\left(#1\right)}}%
\global\long\def\thetat{\theta}%
\global\long\def\thetain{\theta_{i}}%
\global\long\def\p#1{#1{}^{\left(\emph{p}\right)}}%
\global\long\def\n#1{#1{}^{\left(n\right)}}%
\global\long\def\up{\tilde{u}}%
\global\long\def\avgPvec{\left\langle \bm{\mathcal{P}}\right\rangle }%
\global\long\def\m#1{#1{}^{\left(m\right)}}%
\global\long\def\conj#1{#1{}^{*}}%
\global\long\def\M#1{\mathsf{M}_{#1}^{\mathsf{s}}}%
\global\long\def\Ivec#1{\mathsf{I}_{#1}^{\mathsf{s}}}%
\global\long\def\sub#1#2{#1_{#2}}%
\global\long\def\l#1{#1{}^{\left(l\right)}}%
\global\long\def\pvec{\bm{\mathcal{P}}}%
\global\long\def\pvecs#1{\mathcal{P}_{#1}}%
\global\long\def\re{\mathrm{Re}}%
\global\long\def\ensemble#1{\left\langle #1\right\rangle }%
\global\long\def\hom#1{#1^{\left(0\right)}}%
\global\long\def\hpoten{V}%
\global\long\def\cLz{\hom c_{L}}%
\global\long\def\cSz{\hom c_{S}}%
\global\long\def\phase#1{#1^{\left(p\right)}}%
\global\long\def\aa#1{#1{}^{\left(\alpha\right)}}%
\global\long\def\bb#1{#1{}^{\left(\beta\right)}}%
\global\long\def\fmat{\mathsf{F}}%
\global\long\def\amat{\mathsf{A}}%
\global\long\def\pt{\mathcal{PT}}%

\begin{document}

\title{Anomalous energy transport in laminates with exceptional points }
\author{Ben Lustig, Guy Elbaz, Alan Muhafra and Gal Shmuel\thanks{Corresponding author. Tel.: +1 972 778871613. \emph{E-mail address}:
meshmuel@technion.ac.il (G. Shmuel).}\\
 {\small{}{}Faculty of Mechanical Engineering, Technion–Israel Institute
of Technology, Haifa 32000, Israel}}
\maketitle
\begin{abstract}
Recent interest in metamaterials has led to a renewed study of wave
mechanics in different branches of physics. Elastodynamics involves
a special intricacy, owing to a coupling between the volumetric and
shear parts of the elastic waves. Through a study of in-plane waves
traversing  periodic laminates, we here show that this coupling
can result with unusual energy transport.  We find that the corresponding
frequency spectrum contains modes which simultaneously attenuate and
propagate, and demonstrate that these modes coalesce to purely propagating
modes at exceptional points—a property that was recently reported
in  parity-time symmetric systems. We show that the laminate exhibits
metamaterial features near these points, such as negative refraction,
and beam steering and splitting. While negative refraction in laminates
has been demonstrated before by considering pure shear waves impinging
an interface with multiple layers, here we realize it for coupled
waves impinging a simple single-layer interface. This feature, together
with the appearance of exceptional points, are absent from the  model
problem of anti-plane shear waves which have no volumetric part, and
hence from the mathematically identical electromagnetic waves. Our
work further paves the way for applications such as asymmetric mode
switches, by encircling exceptional points in a tangible, purely elastic
apparatus. 

\end{abstract}

\section{Introduction}

Metamaterials possess properties not found in nature, stemming from
their architectured microstructure \citep{Wegener2013,Kadi2019nrp}.
Perhaps the most prominent thrust in metamaterials research aims at
controlling waves for potential applications such as lensing, cloaking
and noise reduction \citep{milton06cloaking,Chen2010nm,PARNELL2013WM,bigoni2013prb,Colquitt2014,Cummer2016yu}.
This interest led to a renewed study of wave mechanics in optics \citep{MarkoSoukoulis_book08,banerjee2011introduction},
acoustics \citep{craster2012acoustic,deymier2013acoustic}, mechanical
lattices \citep{phani2011AIP,raney16,phani2017book,Zelhofer2017kochmann,Ma2018PRL}
and elastodynamics \citep{Brun2010,Shmuel2016JMPS,Chen2017SR,Aghighi2019,LI2019jmps}. 

While electromagnetic-, sound- and anti-plane shear waves are mathematically
identical \citep{adams08,Torrent2011njp}, in-plane elastic waves
are physically richer \citep{Sigalas1992}, since they comprise both
volumetric and distortional parts, coupled through interfaces in the
transmission medium. In this work, we show that in the simplest elastic
composite—a laminate—this coupling gives rise to anomalous energy
transport, which in other systems is achieved by significantly more
complicated means.  The anomalies reported here and their connection
 with recent studies in the field are summarized next.

Firstly, we show that the spectrum of in-plane waves in laminates
exhibits \emph{exceptional points}, accessible in a purely elastic
setting. Exceptional points are states of a system at which two (or
more) of its normal modes coalesce, together with their natural frequencies
\citep{Moiseyev1980PRA,Ding2015PRB}, and are the source of counterintuitive
phenomena such as enhanced sensitivity, wave stopping and asymmetric
transmission \citep{hodaei2017enhanced,Achilleos2017PRB,Goldzak2018PRL,Merkel2018prb}.
Exceptional points occur only in non-Hermitian systems \citep{moiseyev2011book},
a property which usually describes systems that interact with the
environment. The current paradigm\footnote{To break reciprocity, we note that another emerging paradigm is to
employ spatiotemporal composites \citep{Trainiti2016,NASSAR2017jmps,Milton2017} } to access exceptional points is by balancing external gain and loss
through the system, to create parity-time ($\pt$) symmetry \citep{ruter2010observation,shi2016accessing,El-Ganainy2018ys}.
This procedure brings with additional complexity to the elastic medium,
as such realizations require incorporating optomechanical, acoustoelectric
or piezoelectric elements into the system \citep{xu2015PRA,Christensen2016prl,hou2018jap}.
Our findings thus suggest a simpler, purely elastic setting to access
exceptional points, thereby evading these complexities. 

Recently, it was demonstrated that encircling these points leads to
fascinating asymmetric mode switching of microwaves in a metallic
waveguide \citep{Doppler2016nature}. We argue that our setting constitutes
a tangible platform to realize analogous encirclement for elastic
wave switching, having the wavenumber, which is related to the excitation
angle, as the trajectory parameter. As we show in the sequel, this
is made possible owing to the intrinsic Riemann surface structure
of our spectrum near these points in the complex wave vector space.
In contrast with cases where this structure is artificially obtained
by an analytical continuation of an arbitrary parameter (\emph{e.g.},
the frequency in \citealp{SHANIN2018wm}), here, complex wavenumbers
are an intrinsic and accessible part of the spectrum. 

Secondly, we demonstrate that the exceptional points foreshadow anomalous
energy transport, including negative refraction. In this regard,
perhaps our most striking result is the excitation of negative refraction
in laminates by an incoming wave from a homogeneous medium whose interface
with the laminate is parallel to the layers (Fig.$\ $\ref{representative lam hom}).
To put this result into context, we recall that the first report of
negative refraction was for electromagnetic waves, and required a
two-dimensional composite made of complicated split ring resonators
\citep{Smith2000prl,shelby2001experimental,Pendry2004science}. Interest
in this phenomenon has been disseminated to elastodynamics, accompanied
with ongoing studies \citep{craster2012acoustic,CHEN2017JMPS,Bordiga2019apl,NEMATNASSER2019MOM}.
Notably, \citet{willis2013arxivB} was the first to show that a simple
laminate is capable of negatively refracting anti-plane shear waves.
While in the conventional arrangement (interface parallel to the layers),
refraction is always positive, Willis realized that negative refraction
is possible when the interface is normal to the layers (Fig.$\ $\ref{representative lam hom}c).
Based on this interface configuration, further studies of such waves
in laminated media were carried out \citep{Willis2016jmps,NematNasser2015,Srivastava2016jmps,MORINI2019jmps}.
Hence, we demonstrate that in-plane waves may refract negatively in
the conventional arrangement, without the need for this complex apparatus,
nor for gain and loss (\emph{cf}.$\ $\citealp{Hou2018PRApplied}). 

For completeness, we also analyze the transmission problem through
an interface normal to the layers. As highlighted by \citet{Srivastava2017PRA}
and the references therein, a wave incident to such interface induces
an infinite number of transmitted waves, which are required to satisfy
corresponding continuity conditions across the interface. We suggest
a method to calculate the resultant normal mode decomposition owing
to incident in-plane waves, based on suitable orthogonality conditions
\citep{Mokhtari2019arxiv}. Subsequently, we demonstrate analogous
phenomena to those reported by \citet{Srivastava2016jmps}, who studied
the anti-plane problem. These include beam steering—small changes
in the incident angle leading to large changes in the transmission
angle; beam splitting—an incident wave transmitted as simultaneous
negative and positive refracted beams; and pure negative refraction.

Our study is detailed in the forthcomings Secs.$\ $as follows. Sec.$\ $\ref{sec:In-plane-wave-propagation}
firstly revisits the equations governing in-plane waves in infinite
periodic laminates, and subsequently provides two formulations—the
extended plane wave expansion method and the hybrid matrix method—for
their solution \citep{laude2009prb,Tan2010Ultras}.  By analyzing
the pertinent equations, we explain why in contrast with the anti-plane
problem, we here obtain complex wavenumbers, exceptional points, and
negative refraction in the simple configuration. We numerically solve
the eigenvalue problems for an exemplary infinite laminate, and present
its  spectrum in Sec.$\ $\ref{sec:Frequency-Spectrum}. As predicted,
the complex spectrum contains exceptional points with hallmarks of
negative refraction. Sec.$\ $\ref{sec:Transmission-across-an} links
the studied spectrum and eigenmodes to their excitation by incoming
waves from a homogeneous half-space that shares an interface with
the exemplary laminate. Finally, our main results and conclusions
are summarized in Sec.$\:$\ref{sec:Conclusions}.

\section{Equations of in-plane waves in laminated media\label{sec:In-plane-wave-propagation}}

The equations governing in-plane waves in elastic solids can be found
in \citet{graff1975wave}, and their extension to laminated media
appears, \emph{e.g.}, in \citet{Lowe1995} and the references therein
(see also \citealp{Adams2009WRCM} for in-plane Bloch waves in infinitely
 periodic strips). In this Sec., we firstly revisit the equations
for an infinite laminate comprising two alternating layers, and provide
two formulations to determine the resultant normal modes via different
eigenvalue problems. Based on these formulations, we explain next
why in contrast with anti-plane waves, in-plane waves in laminates
admit complex eigenvalues and exceptional points, and connect these
features of the spectrum to the transport of energy.

\subsection{Governing Equations\label{subsec:Hybrid-Matrix-Method}}

\begin{figure}[t]
\includegraphics[width=1\textwidth]{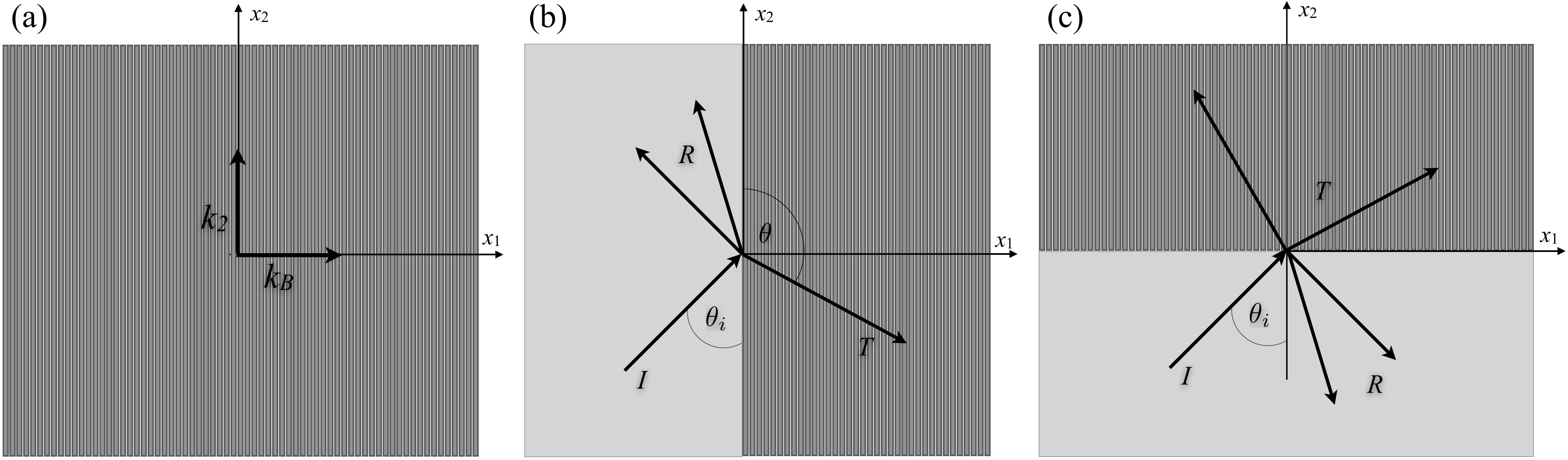}\caption{(a) An infinite laminate comprising alternating $\emph{a}$ and $\emph{b}$
phases. The medium admits Bloch modes with vertical wavenumber $\protect\ksnell$,
and macroscopic horizontal wavenumber $\protect\kb$. (b) A homogeneous
half-space bonded to a semi-infinite laminate at $x_{1}=0$, and at
(c) $x_{2}=0$. We schematically denote the incident, reflected and
transmitted waves  by $I,R$ and $T$, respectively. The incident
and transmission angles are denoted by $\protect\thetain$ and $\theta$,
respectively. }

{\small{}{}\label{representative lam hom}}{\small\par}
\end{figure}

We study an infinite elastic laminate made of a periodic repetition
of phases $\emph{a}$ and $\emph{b}$ in the $x_{1}$ direction (Fig.
\ref{representative lam hom}a). The thickness, mass density and Lamé
coefficients of each phase are $\nthc{\emph{p}},\nrho{\emph{p}},\nlambda{\emph{p}}$
and $\nmu{\emph{p}}$, respectively, where $\emph{p}=\emph{a}$ or
$\emph{b}$, denoting the corresponding phase. 

\floatsetup[figure]{style=plain,subcapbesideposition=top}

The objective is to determine the propagation of free time-harmonic
in-plane waves in the laminate. To this end, we seek solutions for
the displacements $\n{u_{1}}$ and $\n{u_{2}}$ in each $\nth$ layer
using the Naiver-Lamé equations

\begin{equation}
\begin{aligned}\begin{aligned}\left(\p{\lambda}+2\p{\mu}\right)\n{u_{1,11}}+\left(\p{\lambda}+\p{\mu}\right)\n{u_{2,21}}+\p{\mu}\n{u_{1,22}}=\p{\rho}\n{\ddot{u_{1}}},\\
\left(\p{\lambda}+2\p{\mu}\right)\n{u_{2,22}}+\left(\p{\lambda}+\p{\mu}\right)\n{u_{1,12}}+\p{\mu}\n{u_{2,11}}=\p{\rho}\n{\ddot{u_{2}}},
\end{aligned}
\end{aligned}
\label{eq:NL}
\end{equation}
subjected to boundary conditions that will be specified later. We
employ the Helmholtz decomposition to write the in-plane components
of $\n{\ub}$ in terms of scalar potential $\n{\phi}$ and $\n{\psi}$,
namely,

\begin{equation}
\begin{aligned}\n{u_{1}}=\n{\phi_{,1}}+\n{\psi_{,2}},\ \  & \n{u_{2}}=\n{\phi_{,2}}-\n{\psi_{,1}}\end{aligned}
.\label{eq:helmholtz}
\end{equation}
This decomposition simplifies Eq.$\ $\eqref{eq:NL} to the form

\begin{equation}
\begin{aligned}\nabla^{2}\n{\phi}=\frac{1}{\ncLSquare{\emph{p}}}\n{\ddot{\phi}},\ \  & \nabla^{2}\n{\psi}=\frac{1}{\ncSSquare p}\n{\ddot{\psi}},\end{aligned}
\label{eq:waveEq}
\end{equation}
where $\ncL{\emph{p}}=\sqrt{\left(\phase{\lambda}+2\phase{\mu}\right)/\phase{\rho}}$
and $\ncS{\emph{p}}=\sqrt{\phase{\mu}/\phase{\rho}}$ are the pressure
and shear wave velocities of phase $\emph{p}$, respectively. At a
fixed frequency $\omega$, Eq. \eqref{eq:waveEq} is solved by
\begin{equation}
\begin{aligned}\n{\phi}=\ALp{\emph{n}}e^{i\left(\omega t+\kL{\emph{p}}x_{1}-\p{\ksnell}x_{2}\right)}+\ALm{\emph{n}}e^{i\left(\omega t-\kL{\emph{p}}x_{1}-\p{\ksnell}x_{2}\right)},\\
\n{\psi}=\ASp{\emph{n}}e^{i\left(\omega t+\kS px_{1}-\p{\ksnell}x_{2}\right)}+\ASm{\emph{n}}e^{i\left(\omega t-\kS px_{1}-\p{\ksnell}x_{2}\right)},
\end{aligned}
\end{equation}
where 
\begin{equation}
\begin{aligned}\kLSquare{\emph{p}}=\frac{\omega^{2}}{\ncLSquare{\emph{p}}}-\p{\ksnell^{2}},\ \  & \kSSquare{\emph{p}}=\frac{\omega^{2}}{\ncSSquare{\emph{p}}}-\p{\ksnell^{2}},\end{aligned}
\label{eq:klksk2}
\end{equation}
and $\left\{ \n{A_{L/S,\pm}}\right\} $ are integration constants
to be determined from the conditions on the boundaries of layer $n$.
These correspond to the continuity of the traction and displacements
at interfaces between adjacent layers, which immediately requires
$\ksnell^{\left(\emph{a}\right)}=\ksnell^{\left(\emph{b}\right)}\eqqcolon\ksnell$.
The corresponding equations are compactly written in terms of the
state vector $\n{\state}\left(x_{1},x_{2},t\right)$
\begin{equation}
\n{\state}=\left(\n{u_{1}},\n{\sigma_{11}},\n{u_{2}},\n{\sigma_{21}}\right)^{\mathsf{T}};\label{eq:state def new}
\end{equation}
say the boundary between layers $n$ and $n+1$ is at $x_{1}=x_{0}^{\left(n\right)}$,
then the continuity conditions are simply $\n{\state}\left(x_{1}=\n{x_{0}}\right)=\state^{\left(n+1\right)}\left(x_{1}=\n{x_{0}}\right)$.
The coupling between shear and pressure modes enters through the latter
condition, since the components of $\n{\state}$ depend both on $\n{\phi}$
and $\n{\psi}$. The remaining equations stem from the Bloch-Floquet
theorem, which states that over the course of one period the governing
fields are related via the Bloch wavenumber $\kb$, namely,
\begin{equation}
\n{\state}\left(x_{1}=\n{x_{0}}-\n h\right)=e^{i\kb h}\state^{\left(n+1\right)}\left(x_{1}=\n{x_{0}}+\nh{n+1}\right).\label{eq:bloch theorem new}
\end{equation}

\subsection{The hybrid matrix method to determine $\protect\kb\left(\omega\right)$\label{subsec:The-hybrid-matrix}}

Eqs.~\eqref{eq:state def new} and \eqref{eq:bloch theorem new}
can be combined in different ways to deliver an eigenproblem for $\kb$
and $\n{\state}$ as functions of real $\omega$ at prescribed $\ksnell$,
given the laminate composition. The transfer matrix formulation is
the most intuitive and common approach, however it suffers from numerical
instabilities \citep{perez2015relations}. Here, we employ the stable
hybrid matrix method \citep{Tan2010Ultras}, which reads 
\begin{equation}
\left(\begin{array}{ll}
-\Ones 2 & \Hmat_{11}\\
\Zeros 2 & \Hmat_{21}
\end{array}\right)\cdot\sm\left(x_{1},x_{2}\right)=e^{ik_{B}h}\left(\begin{array}{ll}
-\Hmat_{12} & \Zeros 2\\
-\Hmat_{22} & \Ones 2
\end{array}\right)\cdot\sm\left(x_{1},x_{2}\right),\label{eq:ev problem}
\end{equation}
where $\Zeros 2$ (resp. $\Ones 2$) is the $2\times2$ zeros (resp.
unit) matrix, and $\sm$ and $\Hmat_{ij}$ are given in \nameref{Appendix HMM},
together with a detailed derivation of Eq.~\eqref{eq:ev problem}.
The resultant quartic equation for $\kb$ is solved by 
\begin{equation}
\cos\kb h=\frac{-a_{2}\pm\sqrt{a_{2}^{2}-4a_{1}\left(a_{3}-2a_{1}\right)}}{4a_{1}},\label{eq: coskbh}
\end{equation}
where $a_{i}$ are given in \nameref{Appendix HMM}. Eq.$\ $\eqref{eq: coskbh}
provides the \emph{dispersion relation} which relates the microstructure
and mechanical properties of the laminate to the waveform at each
frequency. The structure of Eq.$\ $\eqref{eq: coskbh} implies that
if $\kb$ is a solution, then so are $\pm\kb+2\pi m$ for any $m\in\mathbb{Z}$.
Hence, the real part of all dispersion curves is representable over
the irreducible $1^{\mathrm{st}}$ Brillouin zone\footnote{See, \emph{e.g., }\citet{ZHANG2019jmps} for a discussion on the reciprocal
space symmetries and degeneracies in the general case.} $0<\kb<\pi/h$. The wavenumber $\kb$ can be real, pure imaginary
(henceforth referred as imaginary) \emph{or} complex, as demonstrated
in the sequel, and contrary to the case of anti-plane shear \citep{Willis2016jmps,Srivastava2016jmps}.
This property is essential for the spectrum to exhibit a Riemann structure,
and, in turn, exceptional points. Real $\kb$ corresponds to a propagating
Bloch mode along $x_{1}$, where imaginary $\kb$ corresponds to attenuating
modes, with the exponential decay $e^{-\left|\kb\right|x_{1}}$; since
$e^{in\pi}=\pm1$ for integer $n$, we will also refer to $\kb$ with
Re$\,\kb h=n\pi$ as imaginary, and interpret corresponding modes
as non-propagating. We note that modes with $\kb h=n\pi$ are at the
boundary of the Brillouin zones, and represent standing waves. Complex
$\kb$ describes a progressive mode that exponentially decay according
to Im$\,\kb$. We emphasize that wave attenuation is not an indication
of energy loss, as our system is non-dissipative; it is the result
of a gradual scattering of energy to incoherent waves with zero mean.
Bands of frequencies without real $\kb$ roots are termed directional
band gaps (or simply gaps) since there is a gap in the spectrum in
the $x_{1}$ direction. When there are no real $\ksnell$ solutions
across these bands, they are termed complete gaps, since all directions
of propagation are prohibited.

\subsection{An eigenvalue problem for $\protect\ksnell$ using a plane wave approach
}

The standard plane wave expansion method to obtain the dispersion
relation in elastodynamics dates back to \citet{Sigalas1992} and
\citet{Kushwaha1993}. In photonics, the method has been extended
by \citet{hsue2005prb} to a formulation in which the Bloch wavenumber
is the eigenvalue, and later on by \citet{laude2009prb} for elastodynamics.
A general analysis of the method, associated linear operators, and
the properties of the eigenvalues was carried out recently by \citet{Mokhtari2019arxiv}.
Building upon the approach of \citet{laude2009prb}, we formulate
for our settings an eigenvalue problem in which $\ksnell$ is the
eigenvalue\footnote{We recall that $\ksnell$ is not a Bloch wavenumber and thus is not
identified with any irreducible Brillouin zone (or, conversely, identified
with an infinite one), as the laminate is homogeneous in the $x_{2}$
direction. }. We begin by deriving the standard plane wave method by substituting
into the Cauchy equations of motion
\begin{equation}
\left\{ \left(u_{i,j}+u_{j,i}\right)\mu\left(x_{1}\right)+\lambda\left(x_{1}\right)\delta_{ij}u_{k,k}\right\} _{,j}=\rho\left(x_{1}\right)\ddot{u}_{i},\label{eq:NL EPWE}
\end{equation}
 the Bloch form
\begin{equation}
u_{i}=\up_{i}\left(x_{1}\right)e^{i\left(\omega t-\kb x_{1}-k_{2}x_{2}\right)},\quad\up_{i}\left(x_{1}\right)=\up_{i}\left(x_{1}+h\right),\label{eq: bloch form}
\end{equation}
where here and henceforth, the periodic part of $\left(\circ\right)$
is denoted by $\tilde{\left(\circ\right)}$. Since $\rho\left(x_{1}\right)$,
$\mu\left(x_{1}\right)$, $\lambda\left(x_{1}\right)$ and $\up_{i}\left(x_{1}\right)$
are periodic in $x_{1}$ with a period $h$, they can be written as
\begin{equation}
\prop\left(x_{1}\right)=\sum_{\G}\prop\left(\G\right)e^{-i\G x_{1}},\ \ \ \prop=\rho,\mu,\lambda,\up_{1},\up_{2},\label{eq:prop expansion}
\end{equation}
where $\G=\frac{2\pi m}{h}$, $m\in\mathbb{Z}$, with the  Fourier
coefficients $\left\{ \prop\left(\G\right)\right\} $ 
\begin{equation}
\prop\left(\G\right)=\frac{1}{h}\int_{0}^{h}\prop\left(x_{1}\right)e^{i\G x_{1}}\mathrm{d}x_{1}.
\end{equation}
We have that 

\begin{equation}
\prop\left(\G\right)=\begin{cases}
\frac{\nthc{\emph{a}}}{h}\nprop{\emph{a}}+\frac{\nthc{\emph{b}}}{h}\nprop{\emph{b}}, & \G=0,\\
\left(\pi m\right)^{-1}\left(\nprop{\emph{a}}-\nprop{\emph{b}}\right)\left(1-e^{i\G\nh{\emph{a}}}\right), & \G\neq0.
\end{cases}
\end{equation}
Substituting Eqs.$\ $\eqref{eq: bloch form} and \eqref{eq:prop expansion}
into Eq. \eqref{eq:NL EPWE} and factoring out $e^{i\left(\omega t-\kb x_{1}-k_{2}x_{2}\right)}$
yield \begin{equation}
\begin{split}
& \sum_{\G,\GTag}\left\{ -\ksnell\left[\ksnell \up_{1}\left(\GTag\right)+\left(\GTag+\kb\right)\up_{2}\left(\GTag\right)\right]\mu\left(\G\right)\right.- \\
& -\left.\left(\G+\GTag+\kb\right)\left[\ksnell\lambda\left(\G\right)\up_{2}\left(\GTag\right)+\left(\GTag+\kb\right)\left(\lambda\left(\G\right)+2\mu\left(\G\right)\right)\up_{1}\left(\GTag\right)\right]\right\} e^{-i\left(\G+\GTag\right)x_{1}}= \\
& =-\sum_{\G,\GTag}\omega^{2}\rho\left(\G\right)\up_{1}\left(\GTag\right)e^{-i\left(\G+\GTag\right)x_{1}}, \\
& \sum_{\G,\GTag}\left\{ -\ksnell\left[\left(\GTag+\kb\right)\lambda\left(G\right)+\left(\G+\GTag+\kb\right)\mu\left(\G\right)\right]\up_{1}\left(\GTag\right)-\ksnell^{2}\lambda\left(\G\right)\up_{2}\left(\GTag\right)\right.- \\
& -\left.\mu\left(\G\right)\left[2\ksnell^{2}+\left(\GTag+\kb\right)\left(G+\GTag+\kb\right)\right]\up_{2}\left(\GTag\right)\right\} e^{-i\left(\G+\GTag\right)x_{1}}= \\
& =-\sum_{\G,\GTag}\omega^{2}\rho\left(\G\right)\up_{2}\left(\GTag\right)e^{-i\left(\G+\GTag\right)x_{1}}.
\label{eq: eq motion e1 e2}
\end{split}
\end{equation}We multiply Eqs. \eqref{eq: eq motion e1 e2} by $e^{i\GTTag x_{1}}$
and integrate over one period. Due to Fourier orthogonality, only
terms satisfying $\GTTag=\G+\GTag$ remain, and we end up with an
infinite set of equations that can be cast in matrix form as 
\begin{equation}
\left[\Amat_{0}+\ksnell\Amat_{1}+\ksnell^{2}\Amat_{2}\right]\mathsf{\tilde{u}}\left(\GTag\right)=\omega^{2}\Bmat\mathsf{\tilde{u}}\left(\GTag\right),\label{eq: ev PWE}
\end{equation}
where $\mathsf{\tilde{u}}\left(\GTag\right)$ is a column vector comprising
the Fourier coefficients of $\up_{1}$ and $\up_{2}$, and the matrices
\textbf{$\Amat_{0},\Amat_{1},\Amat_{2}$ }and $\Bmat$ are given in
\nameref{Appendix- EPWE}. Eq.$\ $\eqref{eq: ev PWE} constitutes
a generalized eigenproblem for $\omega^{2}$ and $\mathsf{\tilde{u}}\left(\GTag\right)$
at prescribed $\ksnell$ and $\kb$. A mixed generalized eigenproblem
for $\ksnell$ at prescribed $\omega$ and $\kb$ follows from Eq.$\ $\eqref{eq: ev PWE}
using a state space-like formulation (Chapt.$\ $10 in \citealp{deymier2013acoustic},
see also \citealp{Hussein2014review}), namely,
\begin{equation}
\left[\begin{array}{ll}
\Zeros{} & \Ones{}\\
\omega^{2}\Bmat-\Amat_{0} & -\Amat_{1}
\end{array}\right]\left(\begin{array}{l}
\mathsf{\tilde{u}}\left(\GTag\right)\\
\ksnell\mathsf{\tilde{u}}\left(\GTag\right)
\end{array}\right)=\ksnell\left[\begin{array}{ll}
\Ones{} & \Zeros{}\\
\Zeros{} & \Amat_{2}
\end{array}\right]\left(\begin{array}{l}
\mathsf{\tilde{u}}\left(\GTag\right)\\
\ksnell\mathsf{\tilde{u}}\left(\GTag\right)
\end{array}\right)\label{eq: ev EPWE}
\end{equation}
For completeness, note that an equivalent formulation can be obtained
using the displacement together with the stress and the eigenvector
\citep{Mokhtari2019arxiv}. For computational purposes, the number
of terms in the Fourier series is truncated, say by $-N\leq m\leq N$.
The matrices $\mathsf{A}_{i},\mathsf{B},\mathsf{0}$ and $\Ones{}$
are accordingly of dimension $\left(2N+1\right)\times\left(2N+1\right)$. 

\subsection{\label{subsec:algebric properties}Some properties of $\protect\kb$
and $\protect\ksnell$ and their effect on energy transport }

 To highlight how the coupling between shear and pressure waves affect
$\kb$ and $\ksnell$, we record next their properties in the case
of anti-plane shear in laminates, where such coupling is absent, and
subsequently point out the differences. These differences have significant
implications on frequency spectrum, and in turn the energy flow, since
the slopes of propagating branches are a measure of its mean. More
formally, the average energy flow is 
\begin{equation}
\avgPvec=-\frac{1}{2}\left\langle \mathrm{Re}\left[\sigmab\dot{\ub}^{*}\right]\right\rangle ,
\end{equation}
and satisfies \citep{Willis2016jmps}
\begin{equation}
\avgPvec=\frac{\partial\omega}{\partial\boldsymbol{k}}E,\quad E=\frac{1}{4}\left\langle \sigma_{ij}u_{i,j}^{*}\right\rangle +\frac{1}{4}\omega^{2}\left\langle \rho u_{k}^{*}u_{k}\right\rangle \equiv\frac{1}{2}\omega^{2}\left\langle \rho u_{k}^{*}u_{k}\right\rangle ,\label{eq:Egroupv}
\end{equation}
where $\left\langle \circ\right\rangle $ denotes averaging over one
spatial and temporal period, and $E$ is the total mean energy density;
$\pvec$ is called the acoustic Poynting vector, and the $\boldsymbol{k}$-gradient
of $\omega$ is identified as the group velocity.

\emph{Exceptional points}.—In anti-plane shear, the size of the transfer
matrix is $2\times2$, whose two roots are $\lambda_{1,2}=e^{ik_{B1,B2}h}$.
When the transfer matrix is real, one can show that if $\lambda$
is a solution, then so are $\lambda^{*}$ and $\lambda^{-1}$ . Since
there are only two roots, either $\lambda=\lambda^{*}$ and then $\kb$
is imaginary, or $\lambda^{*}=\lambda^{-1}$ and then $\kb$ is real.
Accordingly, eigenvalue degeneracies occur only at the edge of the
irreducible Brillouin zones, where the modes are standing and not
propagating. In this usual degeneracy, the eigenmodes remain linearly
independent, and the splitting of the eigenvalues from the so-called
\emph{diabolic} \emph{points} scales linearly. 

The situation is significantly different when there are two coupled
displacements as considered here. The size of the corresponding transfer
matrix is $4\times4$,  thus has four roots. Hence, the fact that
$\lambda^{*}$ and $\lambda^{-1}$ are also solutions does not enforce
that $\lambda=\lambda^{*}$ or $\lambda^{*}=\lambda^{-1}$, and, in
turn the exclusion of complex $\kb$ roots. As we show in Sec.~\ref{sec:Frequency-Spectrum},
not only such complex roots exist, they coalesce together with their
eigenmodes at exceptional points \emph{inside} the irreducible $1^{\mathrm{st}}$
Brillouin zone, according to a square-root scaling, constituting a
Riemann surface structure. 

  \emph{Negative refraction}.—We quantify the flow direction of
each mode that is propagating in the laminate plane using the angle
(Fig.$\ $\ref{representative lam hom})

\begin{equation}
\thetat=\arctan\frac{\avgP 1}{\avgP 2}=\arctan\frac{\partial\omega/\partial\kb}{\partial\omega/\partial\ksnell}.
\end{equation}
\citet{Srivastava2016jmps} showed analytically for anti-plane shear
waves that  $\ksnell$ and $\pvecs 2$ share the same sign, and hence
so does $\avgP 2$. Therefore, the macroscopic transport of energy
of anti-plane shear waves in the $x_{2}$ direction is aligned with
the local transport. This implies that in the canonical configuration
of excitation \citep{joseph2015WM},\emph{ i.e.}, when the laminate
is impinged by a wave at a boundary normal to the lamination direction
(Fig. \ref{representative lam hom}b), anti-plane shear will always
refract positively. The reason is that continuity requires the excited
Bloch waves to share the same vertical wavenumber as the incident
wave, and hence the incident angle and the angle of the Bloch waves
have the same sign.  In the present problem, where $\P 2$ is 
\begin{equation}
\P 2=\frac{1}{2}\omega\re\left\{ \mu\left(\ksnell\upm 1+\kb\upm 2+i\dupm 2\right)\conjupm 1+\left(\kb\lambda\upm 1+\ksnell\left(\lambda+2\mu\right)\upm 2+i\lambda\dupm 1\right)\conjupm 2\right\} ,\ \left(\cdot\right)'\equiv\frac{\partial\left(\cdot\right)}{\partial x_{1}},\label{eq:ptexplicit}
\end{equation}
this is no longer the case; as we will demonstrate in the sequel,
there are positive $\ksnell$ for which simultaneously $\ensemble{\pvecs 2}<0$
and $\ensemble{\P 1}>0$, where
\begin{equation}
\P 1=\frac{1}{2}\omega\re\left\{ \mu\left(\ksnell\upm 1+\kb\upm 2+i\dupm 2\right)\conjupm 2+\left(\lambda+2\mu\right)\left(\kb\upm 1+i\dupm 1\right)\conjupm 1+\lambda\ksnell\upm 2\conjupm 1\right\} .\label{eq:poexplicit}
\end{equation}
This implies that negative refraction is realizable in the simple
interface configuration. 

\citet{Willis2016jmps} devised a complex configuration to achieve
negative refraction of anti-plane shear waves, by considering waves
impinging at an interface parallel to the lamination direction (Fig.~\ref{representative lam hom}c).
We will additionally show that in-plane waves can refract negatively
in this excitation setup as well.

\section{Mode spectrum of an exemplary laminate  \label{sec:Frequency-Spectrum}}

In this Sec., we study the frequency spectrum of in-plane waves propagating
through an infinite laminate comprising steel and acrylic layers,
whose properties are
\begin{equation}
\begin{array}{ll}
\nmu 1=78.85\,\mathrm{GPa},\nlambda 1=118.27\,\mathrm{GPa}, & \nrho 1=7800\,\mathrm{kg\,m^{-3}},\nthc 1=1.3\,\mathrm{mm},\\
\nmu 2=1.21\,\mathrm{GPa},\ \nlambda 2=4.86\,\mathrm{GPa}, & \nrho 2=1200\,\mathrm{kg\,m^{-3}},\nthc 2=3\,\mathrm{mm}.
\end{array}\label{eq:lam1}
\end{equation}
The same laminate was considered by \citet{NematNasser2015,Willis2016jmps}
and \citet{Srivastava2016jmps} to study anti-plane shear waves. The
forthcoming study of the spectrum is the basis for the transmission
analysis in Sec.$\ $\ref{sec:Transmission-across-an}. To simplify
the presentation of the spectrum, we fix one of the three parameters
$\left\{ \omega,\ \kb,\ \ksnell\right\} $, and evaluate the relation
between the remaining two. 

\floatsetup[figure]{style=plain,subcapbesideposition=top}

\begin{figure}[t]
\centering\sidesubfloat[]{\includegraphics[width=0.25\textwidth]{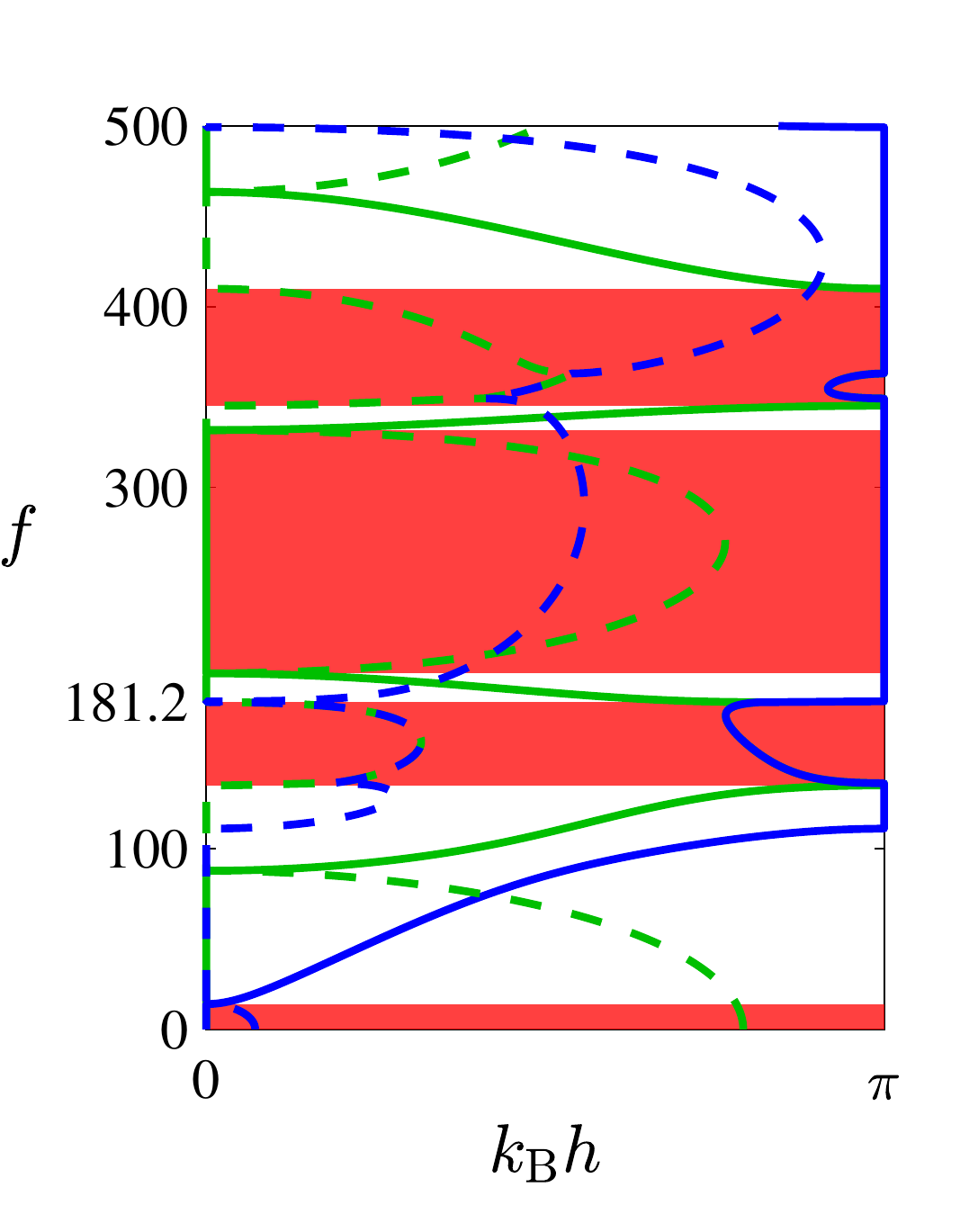}} \quad \sidesubfloat[]{\includegraphics[width=0.25\textwidth]{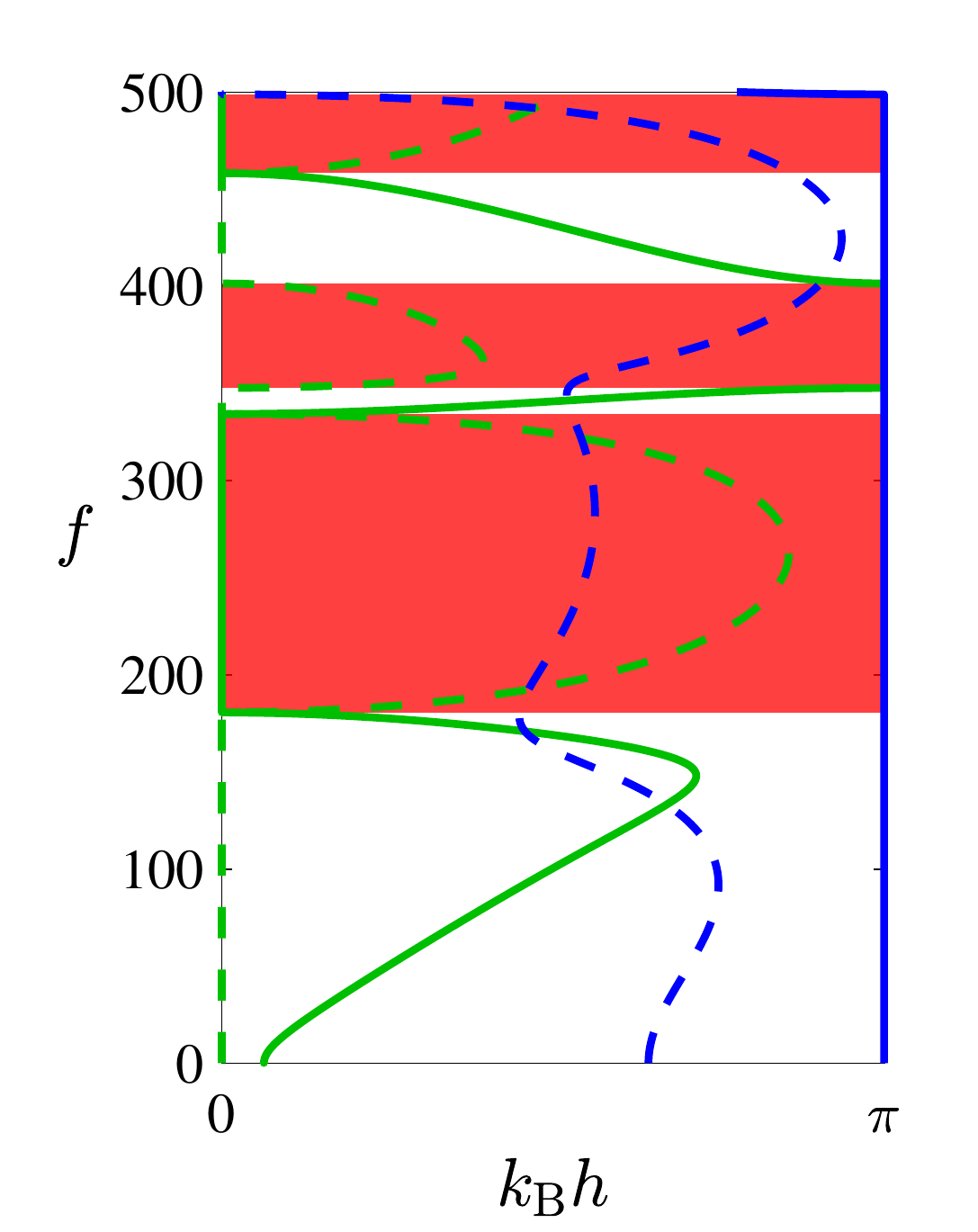}} \quad \sidesubfloat[]{\includegraphics[width=0.25\textwidth]{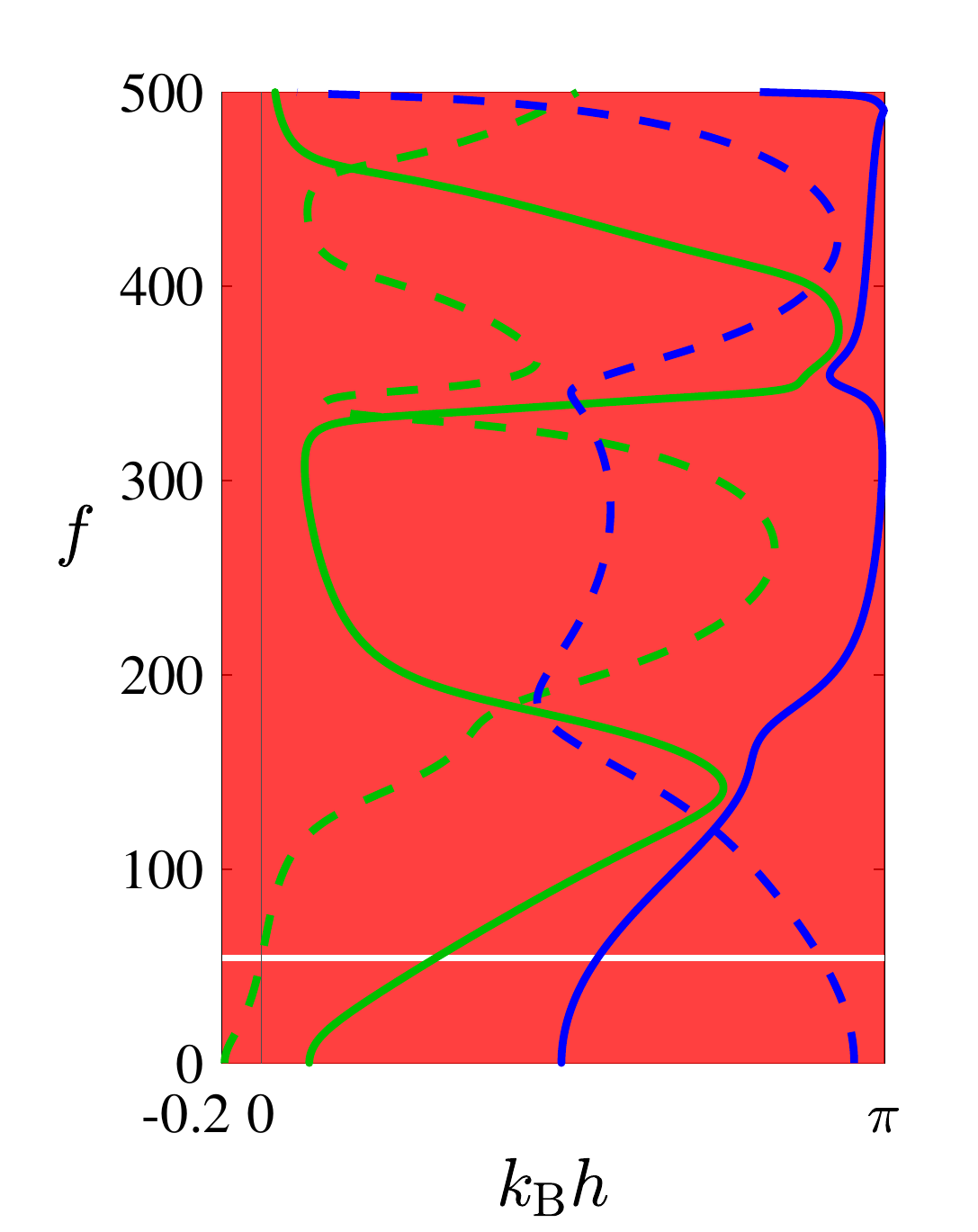}}\caption{The ordinary frequency $f$ (in $\mathrm{kHz}$) versus $\protect\kb h$
restricted to $0<\protect\kb h<\pi$ for laminate \eqref{eq:lam1}
at (a) $\protect\ksnell h=0.5$, (b) $0.5i$, and (c) $0.5+0.5i$.
Re$\,\protect\kb h$ and |Im$\,\protect\kb h$| (Im$\,\protect\kb$
in panel c) are shown in solid and dashed curves, respectively. The
real and imaginary parts of a certain branch are plotted using the
same color. Gaps are indicated by the red shading.}

{\small{}{}\label{fig omega against kb}}{\small\par}
\end{figure}
 Our analysis begins by plotting in Fig.~\ref{fig omega against kb}
the ordinary frequency $f=\omega/2\pi$ versus $0<\kb h<\pi$ at prescribed
$\ksnell$ values. Panels \ref{fig omega against kb}(a)-(c) correspond
to $\ksnell h=0.5,\ 0.5i,$ and $0.5+0.5i$, respectively. Re$\,\kb h$
and |Im$\,\kb h$| (Im$\,\kb h$ in panel c) are shown in solid and
dashed curves, respectively, where the real and imaginary parts of
a certain branch are plotted using the same color. Gaps are indicated
by the red shading.

Panel \ref{fig omega against kb}(a) exhibits the following notable
features. Firstly, there are no branches with real $\kb$ from $f=0$
to $14.1\,\mathrm{kHz}$.  In a study of the dependency of this gap
on Re$\,\ksnell h$ (not shown here), we found that it widens as Re$\,\ksnell h$
increases. Across $136.2<f<181.2\ \mathrm{kHz}$, the branches are
complex conjugates of each other; from $f=159.3\ \mathrm{kHz}$ the
imaginary parts decrease until they vanish at $f\approx181.2\ \mathrm{kHz}$.
At this exceptional point the eigenvalues coalesce. Beyond this point
there is a special narrow range where both branches are real, while
their slopes have an opposite sign, \emph{i.e.}, the modes are propagating
in opposite directions; this is the fingerprint of exceptional points. 

Panel \ref{fig omega against kb}(b) exhibits different characteristics.
Firstly, across the studied frequencies, one of the branches always
has an imaginary part, hence the maximal number of modes that propagate
in $x_{1}$ at any frequency is one. Contrary to the case in panel
\ref{fig omega against kb}(a), here the first gap emerges above $f=0$,
namely, at $f=180.6\ \mathrm{kHz}$. In a study of the dependency
of this gap on Im$\,\ksnell h$ (not shown here), we found that gaps
starting at $f=0$ emerge at higher values of Im$\,\ksnell h$. Thus,
the diagram starts with a propagating band, where the horizontal group
velocity changes sign at $f=148\ \mathrm{kHz}$. This flip of sign
\emph{inside }the irreducible first Brillouin zone is unique to in-plane
waves. 

Panel \ref{fig omega against kb}(c) shows an anomalous scenario without
passbands, except at a discrete frequency ($f\approx54.2\,$kHz) for
which one branch is propagating. To facilitate the visual identification
of this frequency, here we plot the signed imaginary part instead
of its absolute value. Accordingly, this frequency is spotted  by
the zero crossing of Im$\,\kb h$. There is no counterpart to this
phenomenon in anti-plane waves, where complex $\ksnell$ are not accessible.

\floatsetup[figure]{style=plain,subcapbesideposition=top}

\begin{figure}[t]
\centering\sidesubfloat[]{\includegraphics[width=0.45\textwidth]{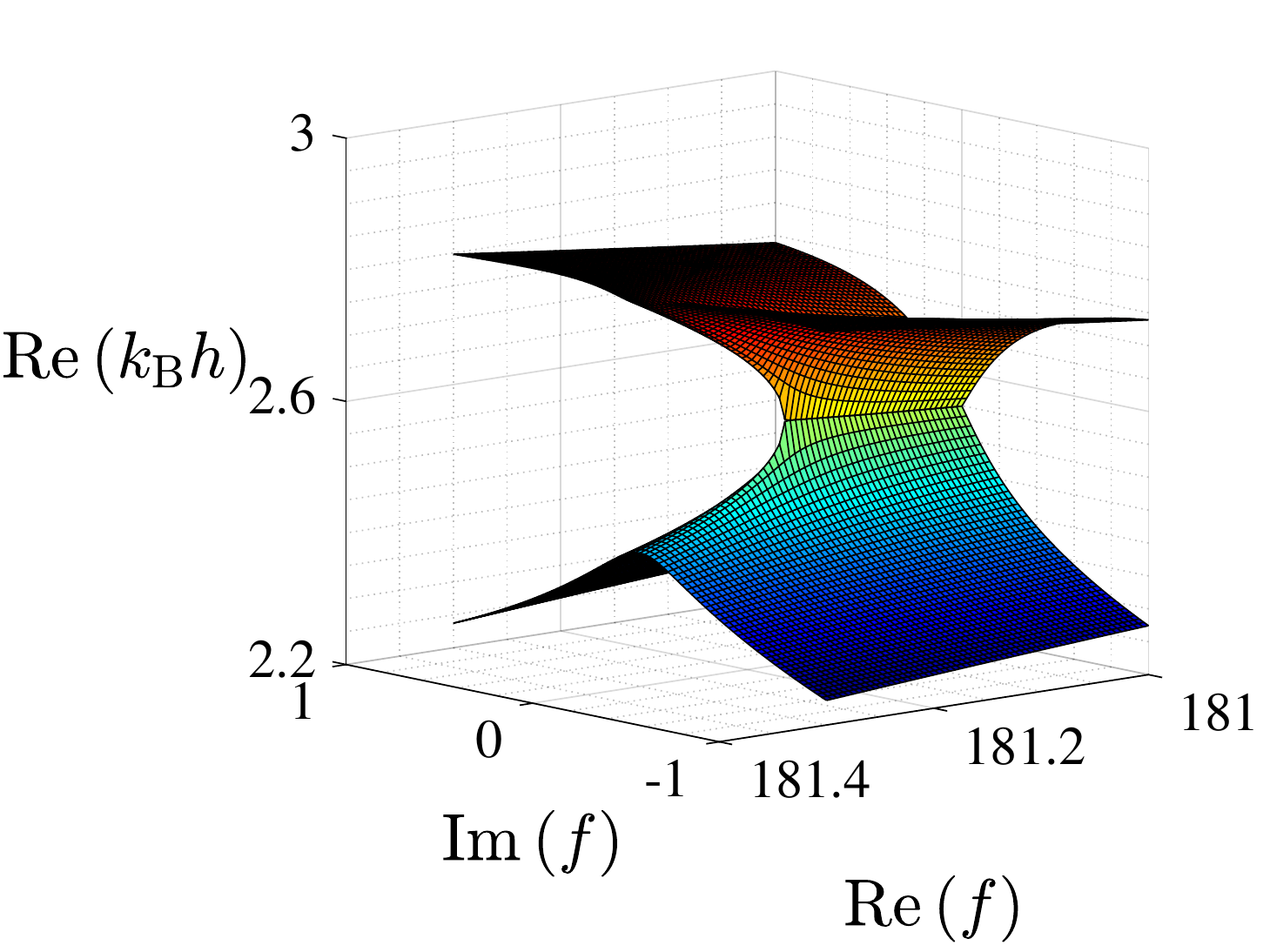}} \quad \sidesubfloat[]{\includegraphics[width=0.45\textwidth]{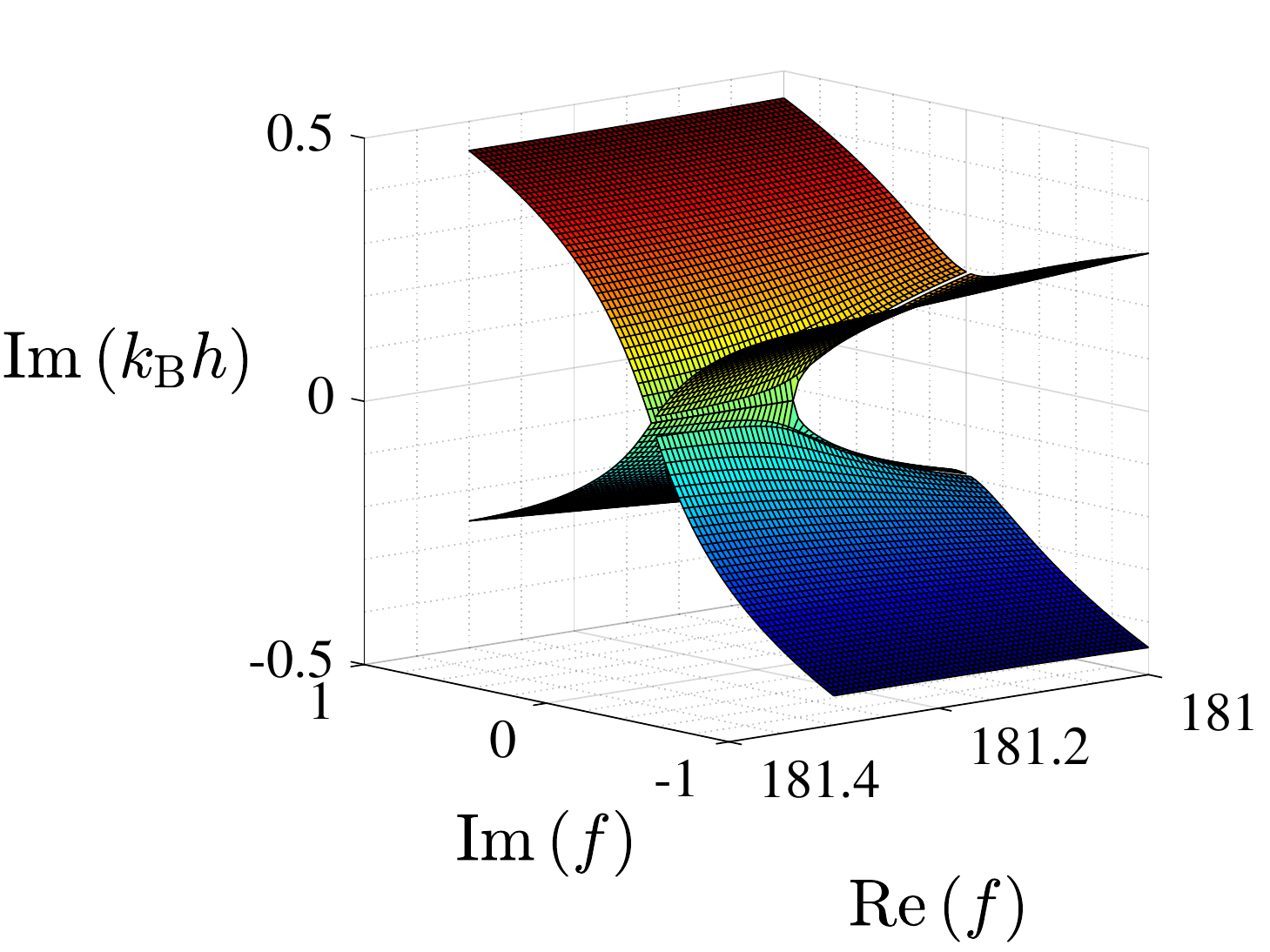}}\caption{The (a) real and (b) imaginary part of $\protect\kb h$ versus the
real and imaginary parts of $f$, for $\protect\ksnell h=0.5$.}

{\small{}{}\label{fig exceptional point f plane}}{\small\par}
\end{figure}
Fig.~\ref{fig exceptional point f plane} shows the (a) real and
(b) imaginary parts of $\kb h$ as functions of $f$ at $\ksnell h=0.5$,
when we carry out an analytic continuation to the dispersion relation,
such that the domain of the spectrum is formally extended to complex-valued
frequencies. We emphasize that here, this is an artificial continuation
(see, \emph{e.g.}, \citealp{SHANIN2018wm}), as complex frequencies
are not accessible (we are concerned with time-harmonic waves in a
system with no dissipation). \citet{lu2018level}, for example, considered
an analytic continuation of the shear modulus, which has the physical
interpretation of a system with gain or loss.

The spectrum exhibits a structure of a Riemann surface in the vicinity
of the exceptional point: Fig.~\ref{fig omega against kb}(a) is
thus a section of that surface, at  the plane Im$\,f=0$. In view
of the duality of between the frequency and the wavenumbers \citep{Torrent2018prb}
in the different forms of the eigenvalue problem \citep{Mokhtari2019arxiv},
a Riemann surface is expected when $f$ is fixed and the roots of
$\kb$ are evaluated against $\ksnell$, as we will demonstrate in
the sequel. Importantly, the states of the system in the complex wave
vector space are accessible, and are not merely a formal extension.

\floatsetup[figure]{style=plain,subcapbesideposition=top}

\begin{figure}[t]
\centering\sidesubfloat[]{\includegraphics[width=0.45\textwidth]{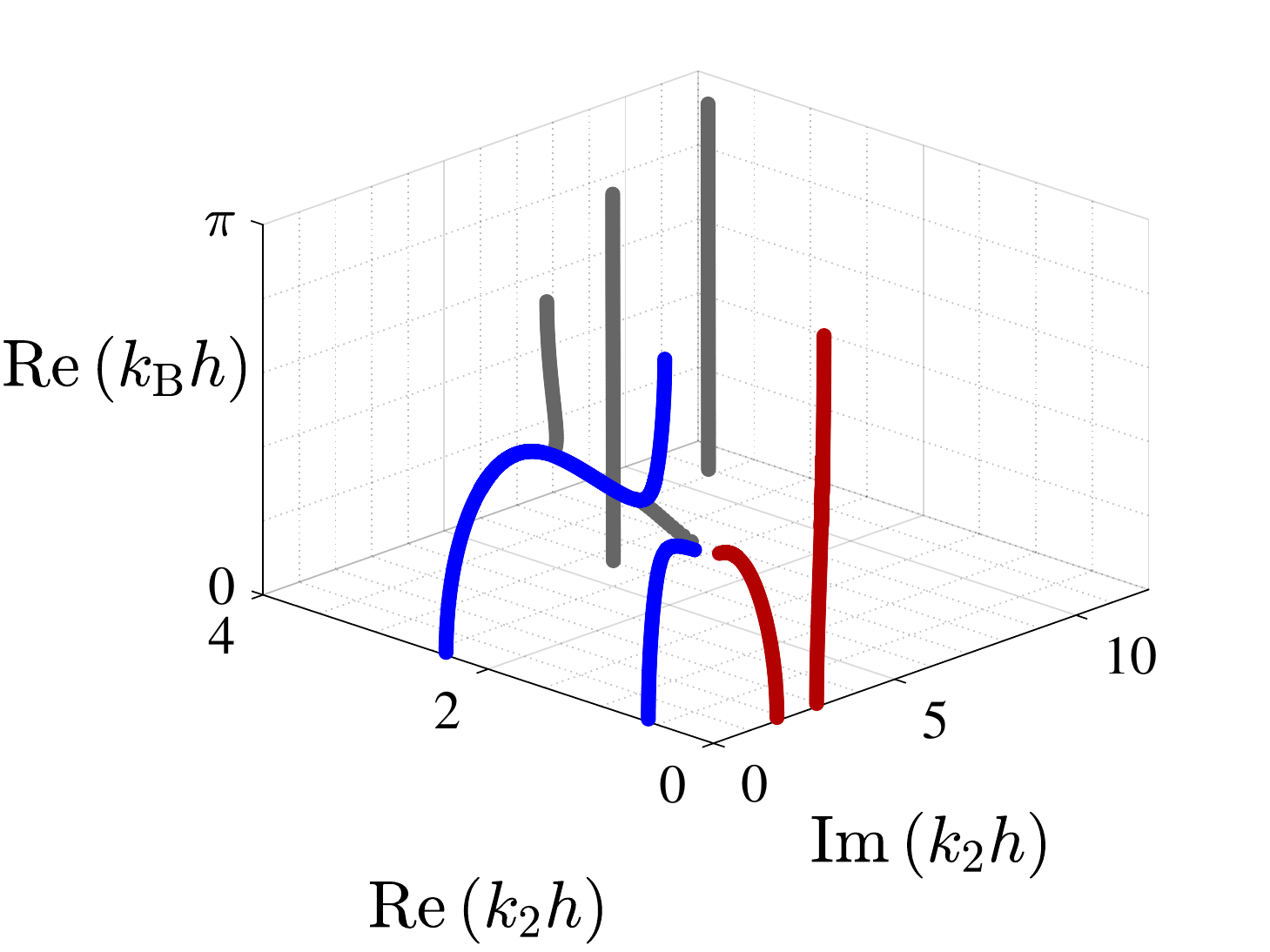}} \quad \sidesubfloat[]{\includegraphics[width=0.45\textwidth]{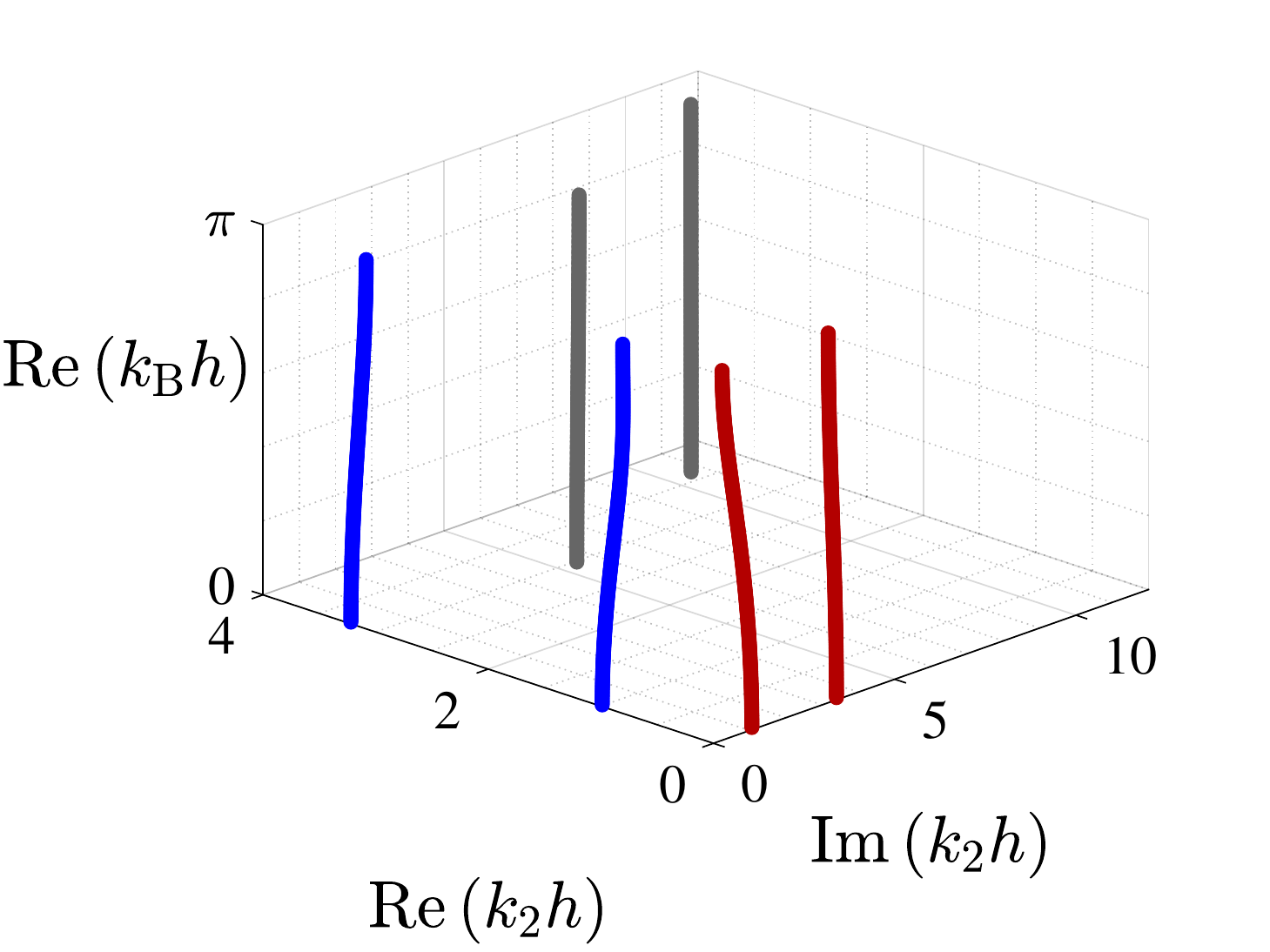}}\caption{Pure real $\protect\kb h$ branches as function of the real and imaginary
parts of $\protect\ksnell h$ for laminate \eqref{eq:lam1} at (a)
$f=100\ \mathrm{kHz}$, and (b) $160\ \mathrm{kHz}$. Segments of
real, imaginary and complex $\protect\ksnell$ are denoted by blue,
red, and grey, respectively. }

{\small{}{}\label{fig kb against k2}}{\small\par}
\end{figure}
Towards this end, we fix $f$ and examine the spectrum in the complex
 $\kb-\ksnell$ space. Figs.~\ref{fig kb against k2}(a) and \ref{fig kb against k2}(b)
show branches of purely real $\kb$ versus Re$\,\ksnell h$ and Im$\,\ksnell h$,
at $f=100\ \mathrm{kHz}$ and $160\ \mathrm{kHz}$, respectively\footnote{The diagrams were evaluated using the extended plane wave method with
51 plane waves in the expansion. A comparison with the exact hybrid
matrix method can be found in \nameref{Appendix-COMPARISON EPWE AND HMM}.}. Segments of real, imaginary and complex $\ksnell$ are denoted by
blue, red, and grey, respectively. Notably, the number of modes with
real or imaginary $\ksnell$ is finite. In our study (not shown here)
the number of modes with real or imaginary $\ksnell$ increases with
the frequency. We further note that in our study on branches with
complex $\ksnell$, we found that their number is infinite, although
it cannot be observed from the truncated diagram we show. By contrast,
\citet{Srivastava2016jmps} showed that in the anti-plane motion there
are no complex $\ksnell$, and the number of branches with \emph{imaginary}
$\ksnell$ is infinite. In both cases (anti-plane and in-plane waves),
the existence of an infinite number of decaying modes conforms with
the need of such a set in  satisfying the continuity of the displacement
and traction across certain interfaces \citep{Srivastava2017PRA}.
\floatsetup[figure]{style=plain,subcapbesideposition=top}

\begin{figure}[t]
\centering\sidesubfloat[]{\includegraphics[width=0.28\textwidth]{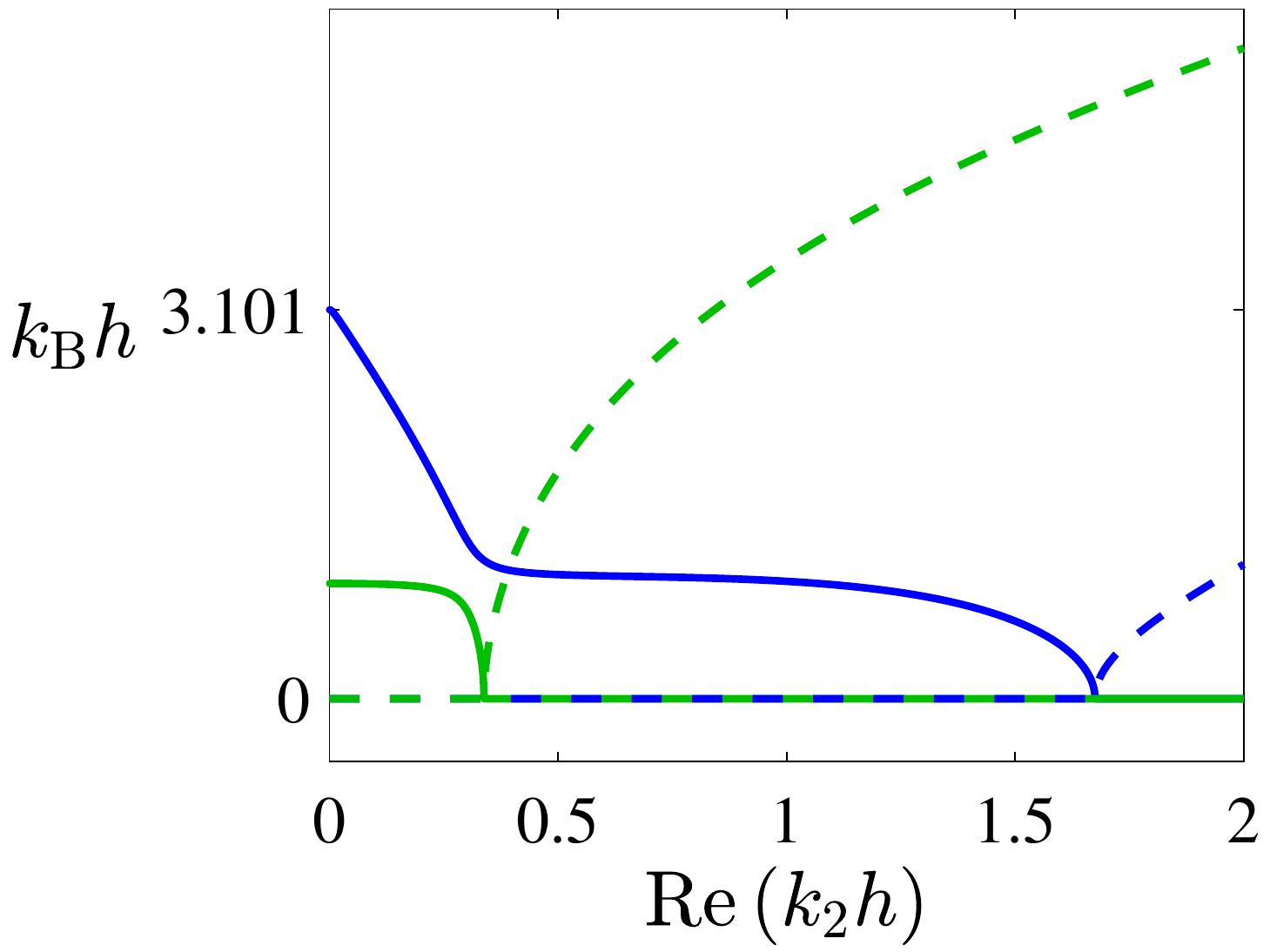}} \quad \sidesubfloat[]{\includegraphics[width=0.28\textwidth]{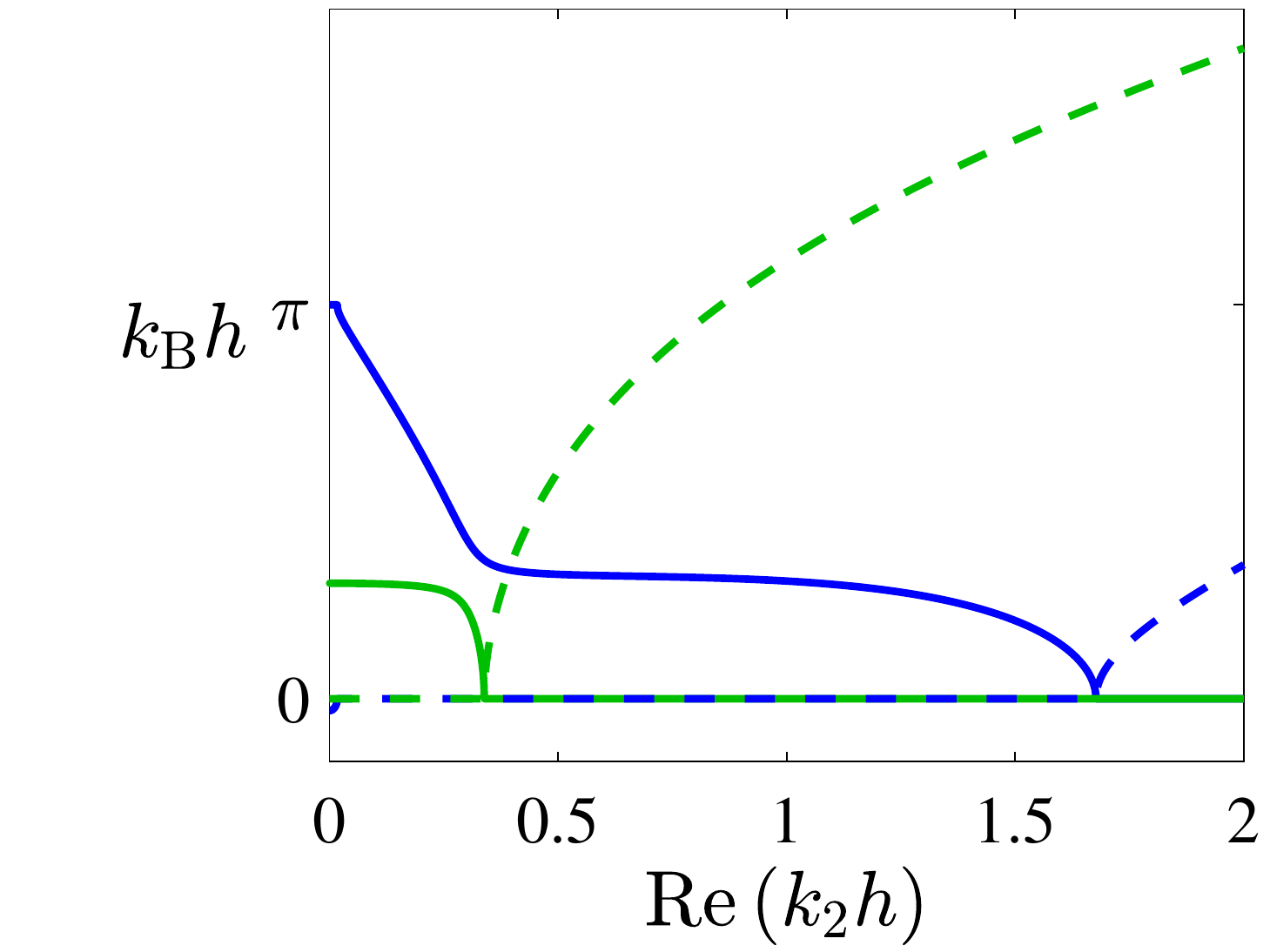}} \quad \sidesubfloat[]{\includegraphics[width=0.28\textwidth]{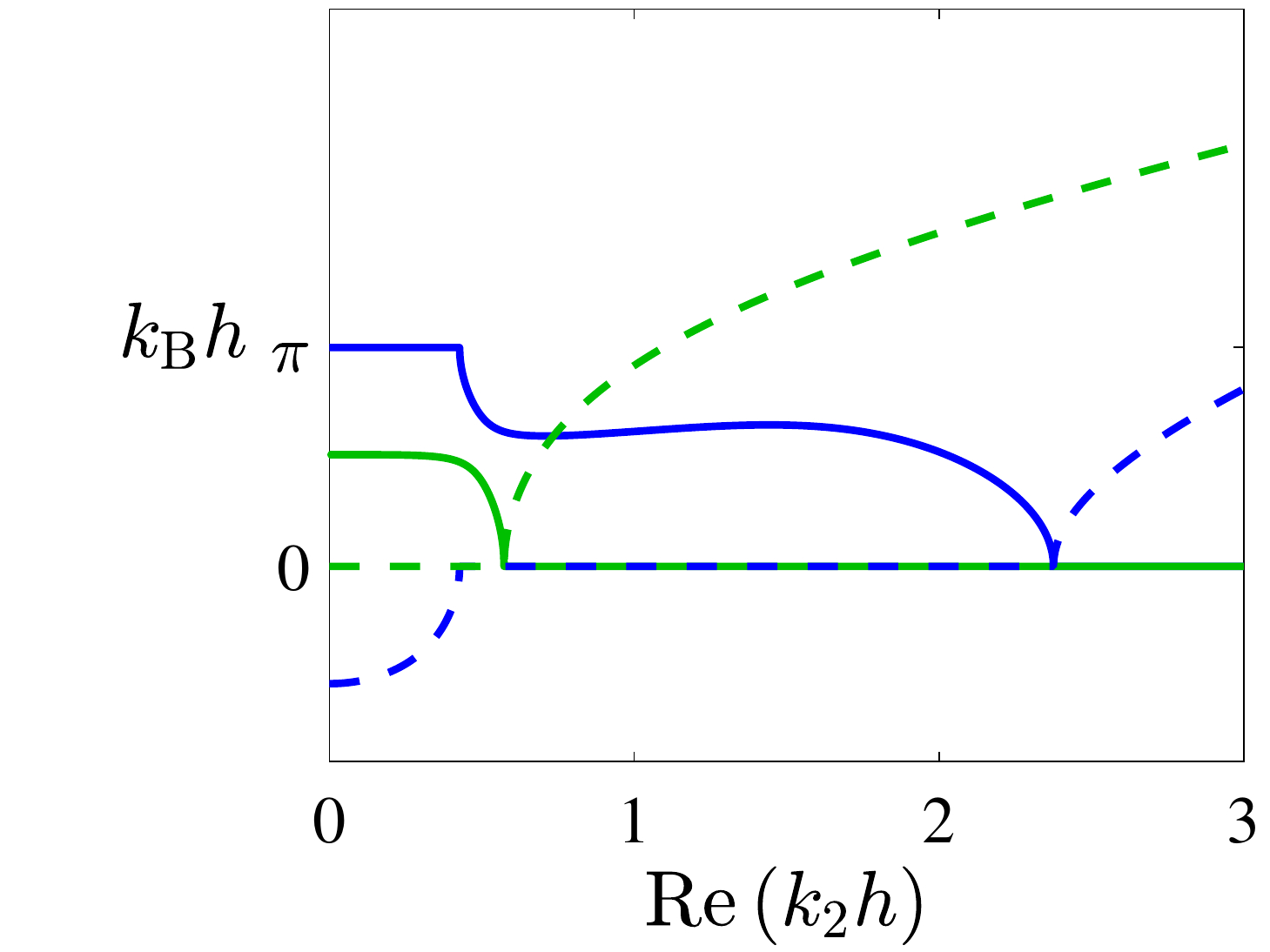}} \quad \sidesubfloat[]{\includegraphics[width=0.28\textwidth]{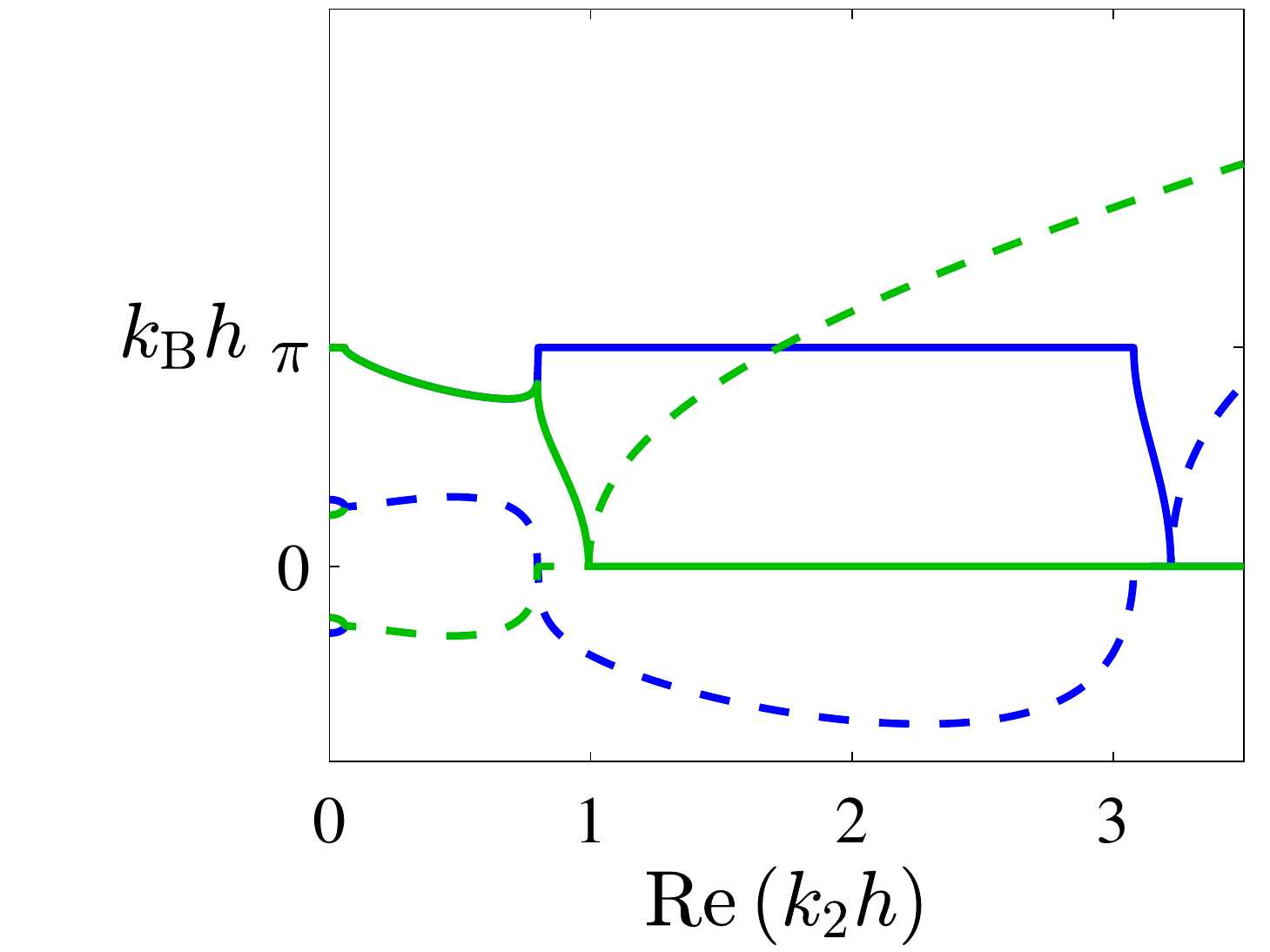}} \quad \sidesubfloat[]{\includegraphics[width=0.28\textwidth]{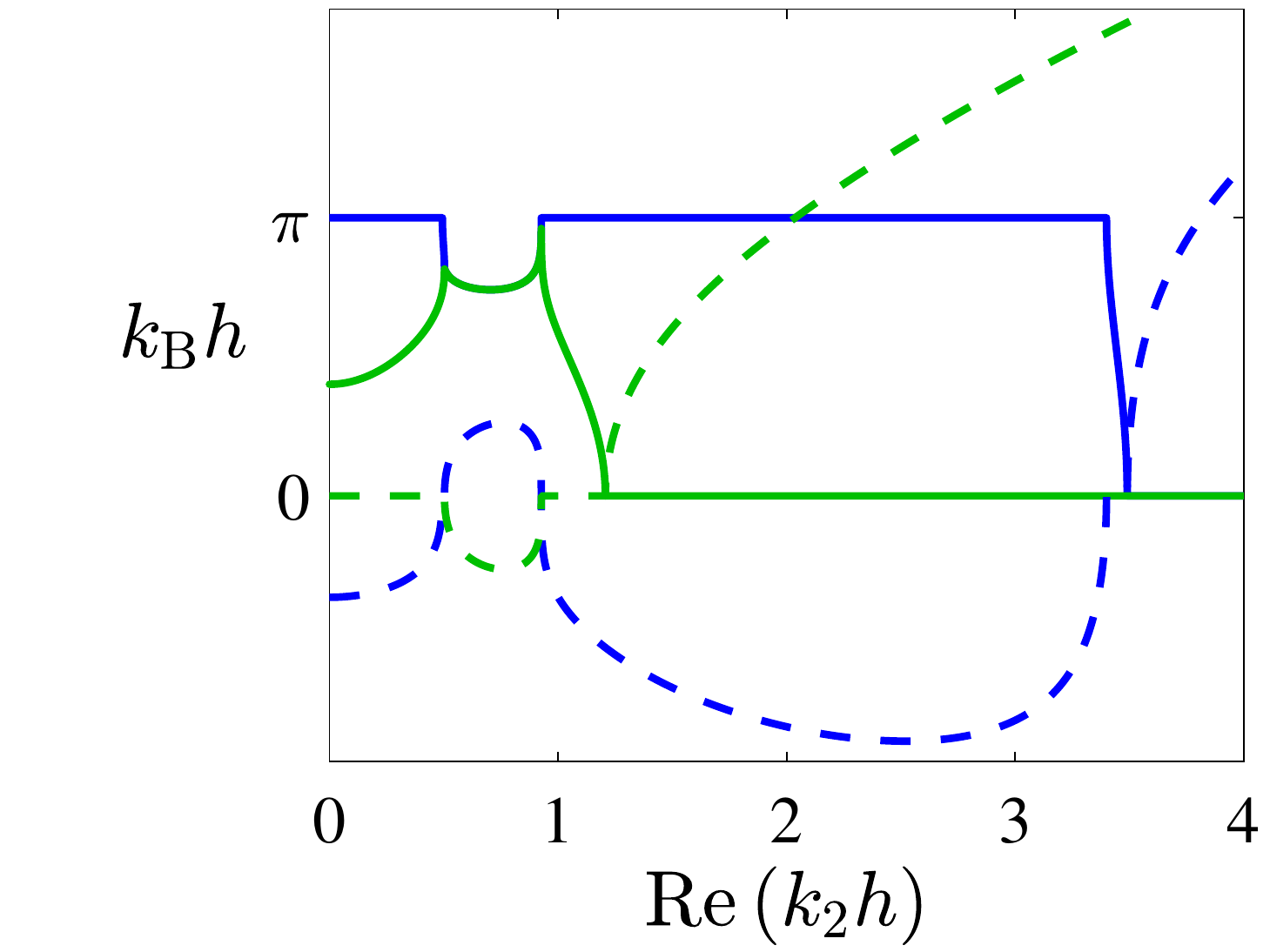}} \quad \sidesubfloat[]{\includegraphics[width=0.28\textwidth]{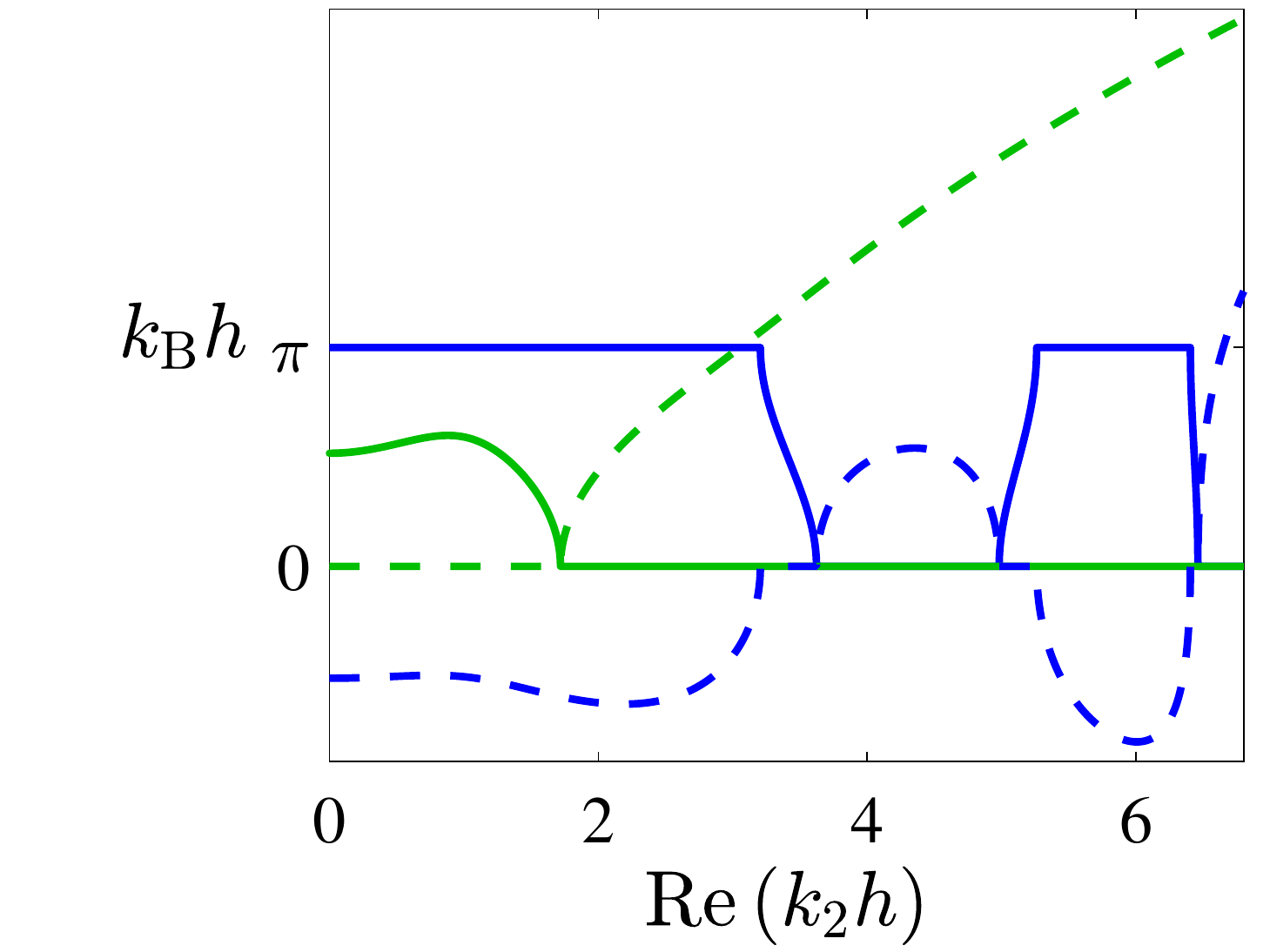}}\caption{The real (solid curves) and imaginary (dashed curves) parts of $\protect\kb h$
as functions of Re\textbf{$\,\protect\ksnell h$} for laminate \eqref{eq:lam1},
at Im$\,\protect\ksnell h=0$ and (a) $f=60$, (b) $60.1$, (c) $100$,
(d) $160$, (e) $181.3$, and (f) $340\,\mathrm{kHz}$. Parts of the
same branch are plotted using the same color. }

{\small{}{}\label{fig prop modes}}{\small\par}
\end{figure}
Modes that propagate along both $x_{1}$ and $x_{2}$ are studied
in Fig. \ref{fig prop modes}, by restricting attention to the plane
Im$\,\ksnell h=0$, and evaluating $\kb h$ versus Re$\,\ksnell h$
at prescribed frequencies. The real and imaginary parts of $\kb$
are shown in solid and dashed curves, respectively, where each color
corresponds to a different branch.

At $f=60\ \mathrm{kHz}$ (panel \ref{fig prop modes}a), two modes
of different branches are propagating when $0<\kb h<0.919$; only
one mode propagates when $0.919<\kb h<3.101$, and at larger Bloch
wavenumbers there are no propagating modes. Here and at higher frequencies,
we find that beyond the illustrated $\ksnell$-interval there are
no propagating branches---solutions at higher $\ksnell$ values are
always given by $\kb$ with an imaginary part, such that $\kb$ is
either imaginary or complex.

At a slightly higher frequency ($f=60.1\ \mathrm{kHz}$, panel \ref{fig prop modes}b),
the whole $1^{\mathrm{st}}$ Brillouin zone admits real solutions,
contrary to the case at $f=60\ \mathrm{kHz}$. This change in the
spectrum between the frequencies has a significant implication on
beam steering, as we will show in Sec.$\,$\ref{subsec:tangetboundary}.
In a narrow range near $\ksnell h=0$, the blue mode has a non-zero
imaginary part and its real part equals $\pi$. Thus, for the frequencies
in panels \ref{fig prop modes}(a)-(b) and displayed range of $\ksnell h$
values, the Bloch wavenumber of the modes is either real, or has an
imaginary part with Re$\,\kb h=0$ mod $\pi$.

Panel \ref{fig prop modes}(c), which evaluates the modes at $f=100\ \mathrm{kHz}$,
also shares the latter characteristic. Here, however, the real part
of the blue is a non-monotonic function of $\ksnell h$, as it changes
trend twice. This is the fingerprint of an exceptional point, and
an indicator of negative refraction as will be demonstrated in Sec.~\ref{subsec:tangetboundary}.
Similarly to the observation made regarding the modes near the exceptional
point in Fig.~\ref{fig omega against kb}, the slopes of the two
modes have different sign.  There are accordingly three propagating
modes across the range $1.87\apprle\kb h\lesssim2.03$ associated
with the same branch, which exhibit different vertical wavelengths.
These features are unique to the in-plane motion, as in the anti-plane
motion the real part of $\kb h$ is either monotonically increasing
to $\pi$ or monotonically decreasing to $0$ \citep{Srivastava2016jmps}.

At $f=160\ \mathrm{kHz}$ (panel \ref{fig prop modes}d) across the
interval $0<\mathrm{Re}\,\ksnell h\lesssim0.79$, there are two complex
conjugate pairs of $\kb$ solutions with and identical real part,
 up to $\ksnell h\approx0.79$.  At this exceptional point, the
imaginary part of all the branches vanishes,  and the modes coalesce.
Beyond this point, the real part of the branches diverges in a different
directions, similarly to the feature observed in Fig. \ref{fig omega against kb}(a).
At $\ksnell h\approx0.8$, the blue branch becomes attenuating, and
at $\ksnell h\approx1$ the green branch also becomes imaginary. The
blue mode is propagating again for $3.08<\ksnell h<3.22$; beyond
that range there no more propagating modes.

When $f=181.3\,\mathrm{kHz}$ (panel \ref{fig prop modes}e), there
are two exceptional points at $\ksnell h=0.503$ and $0.93$. Again,
in the vicinity of these points there are modes with slopes of different
sign. As in panel \ref{fig omega against kb}(d), there is a range
of Bloch wavenumbers with three propagating modes when $\kb h>1.26$,
where for lower $\kb h$ there are two propagating modes. 

We recall that in Fig.$\ $\ref{fig omega against kb}(a), we exhibited
an exceptional point for $f=181.2\,\mathrm{kHz}$ and $\ksnell h=0.5$,
\emph{i.e.}, when perturbing one of the parameters of the system about
this point (\emph{e.g.}, $\ksnell h$ is perturbed from 0.5 to 0.503),
another exceptional point emerges at a variation of another parameter
of the system (\emph{e.g.}, is perturbed $f$ from 181.2 to $181.3\,\mathrm{kHz}$).
This is not accidental; in fact, the exceptional points we demonstrate
for each set of parameters comprise \emph{exceptional curves} in a
higher dimensional space.

At $f=340\,\mathrm{kHz}$ (panel \ref{fig prop modes}f), the propagating
green mode is a non-monotonic function of $\ksnell$, where initially
it has a positive slope up to a maximal point, beyond which the slope
changes sign, until the mode becomes imaginary at $\ksnell h=1.72$.
The blue mode is either real or imaginary with $\mathrm{Re}\,\kb h$
mod $\pi=0$. The number of propagating modes for prescribed $\kb$
is higher than at lower frequencies, namely, a minimal number of three
and a maximal number of five across $1.624<\kb h<1.88$.\floatsetup[figure]{style=plain,subcapbesideposition=top}

\begin{figure}[t]
\centering\sidesubfloat[]{\includegraphics[width=0.45\textwidth]{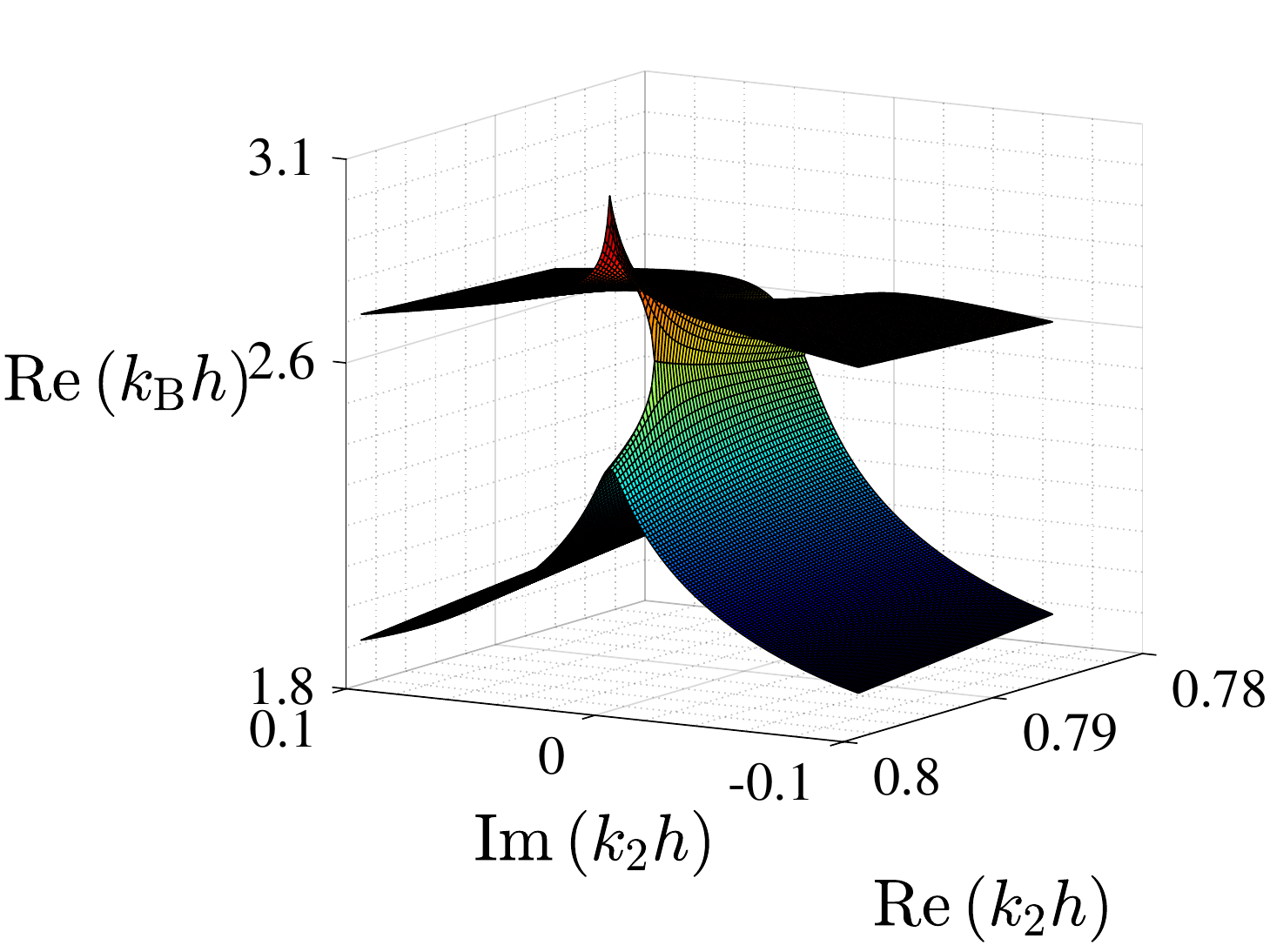}} \quad \sidesubfloat[]{\includegraphics[width=0.45\textwidth]{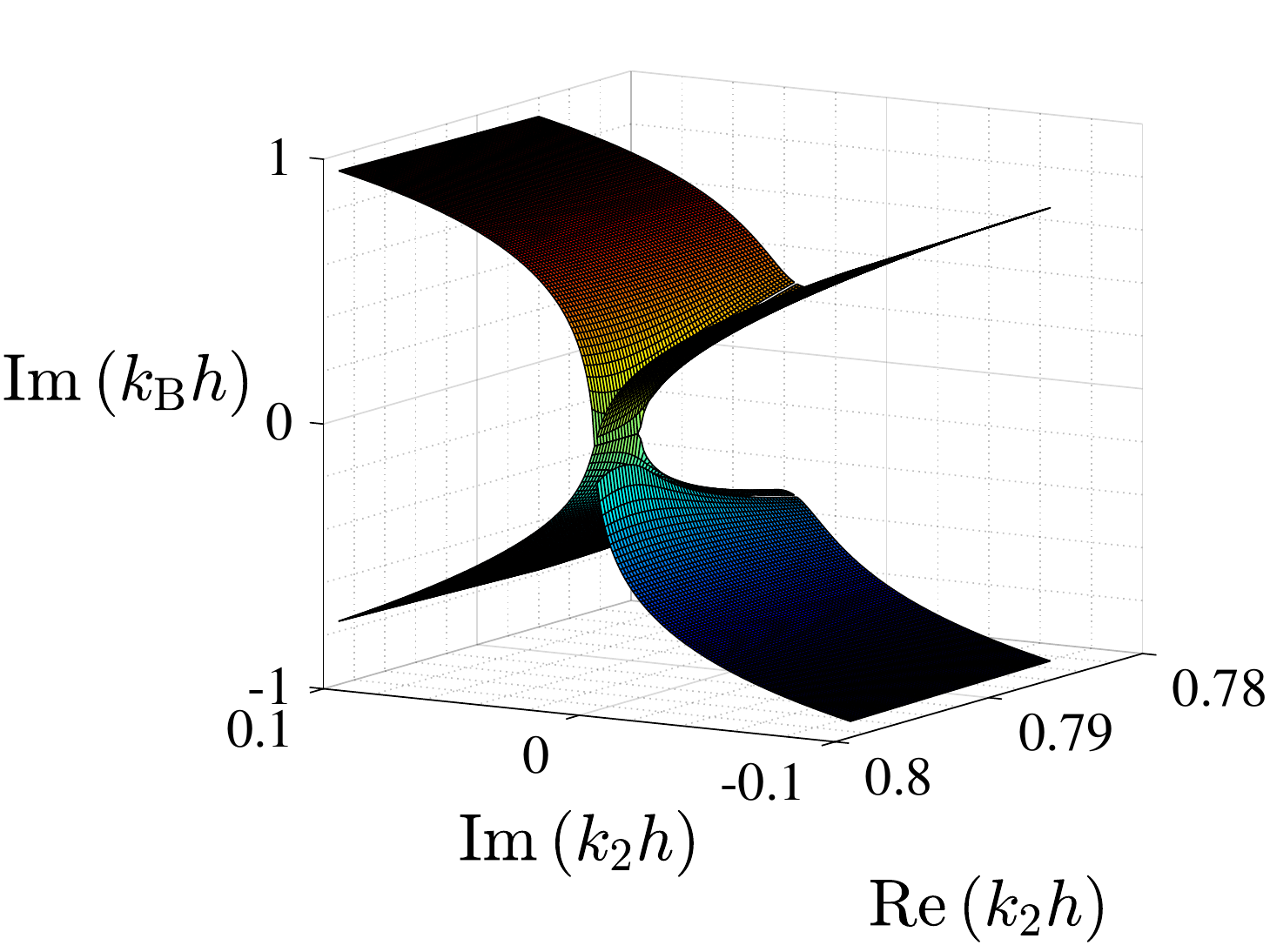}}\caption{The (a) real and (b) imaginary parts of $\protect\kb h$ as functions
of Re$\,\protect\ksnell h$ and Im$\,\protect\ksnell h$, at $f=160\,$kHz.}

{\small{}{}\label{fig exceptional point k2h plane}}{\small\par}
\end{figure}
As mentioned earlier, we expect the spectrum to exhibit the structure
of a Riemann surface near exceptional points in the complex $\kb-\ksnell$
space. This is shown in Fig.~\ref{fig exceptional point k2h plane},
where the (a) real and (b) imaginary parts of $\kb h$ are plotted
against Re$\,\ksnell h$ and Im$\,\ksnell h$, at $f=160\,$kHz:
Fig.~\ref{fig prop modes}(d) is thus the section Im$\,\ksnell=0$
of this surface. We emphasize again the the states of the system over
this manifold are accessible since they correspond to realizable wavenumbers,
and are not merely a formal extension.

\floatsetup[figure]{style=plain,subcapbesideposition=top}

\begin{figure}[t]
\centering\sidesubfloat[]{\includegraphics[width=0.25\textwidth]{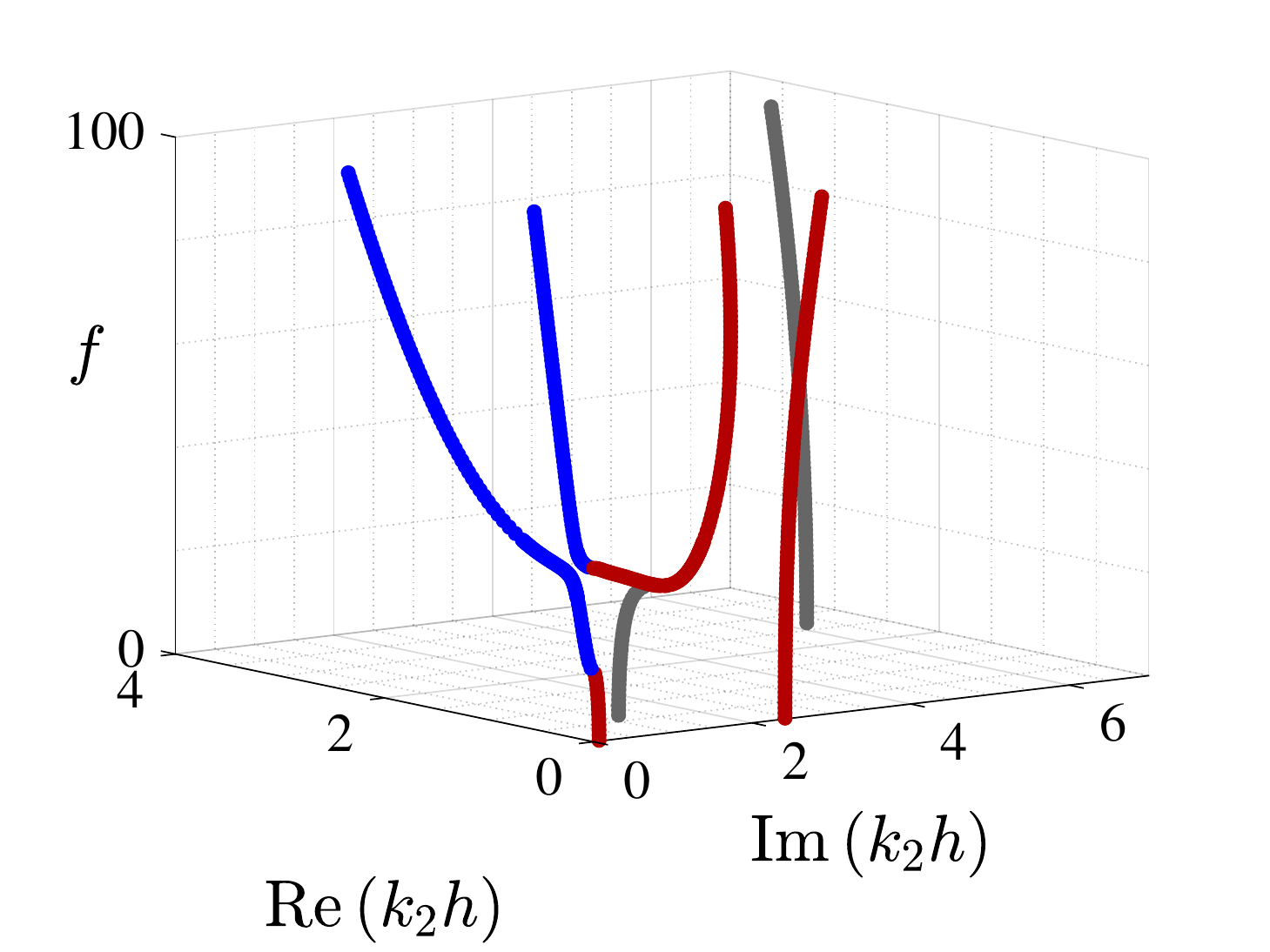}} \quad \sidesubfloat[]{\includegraphics[width=0.25\textwidth]{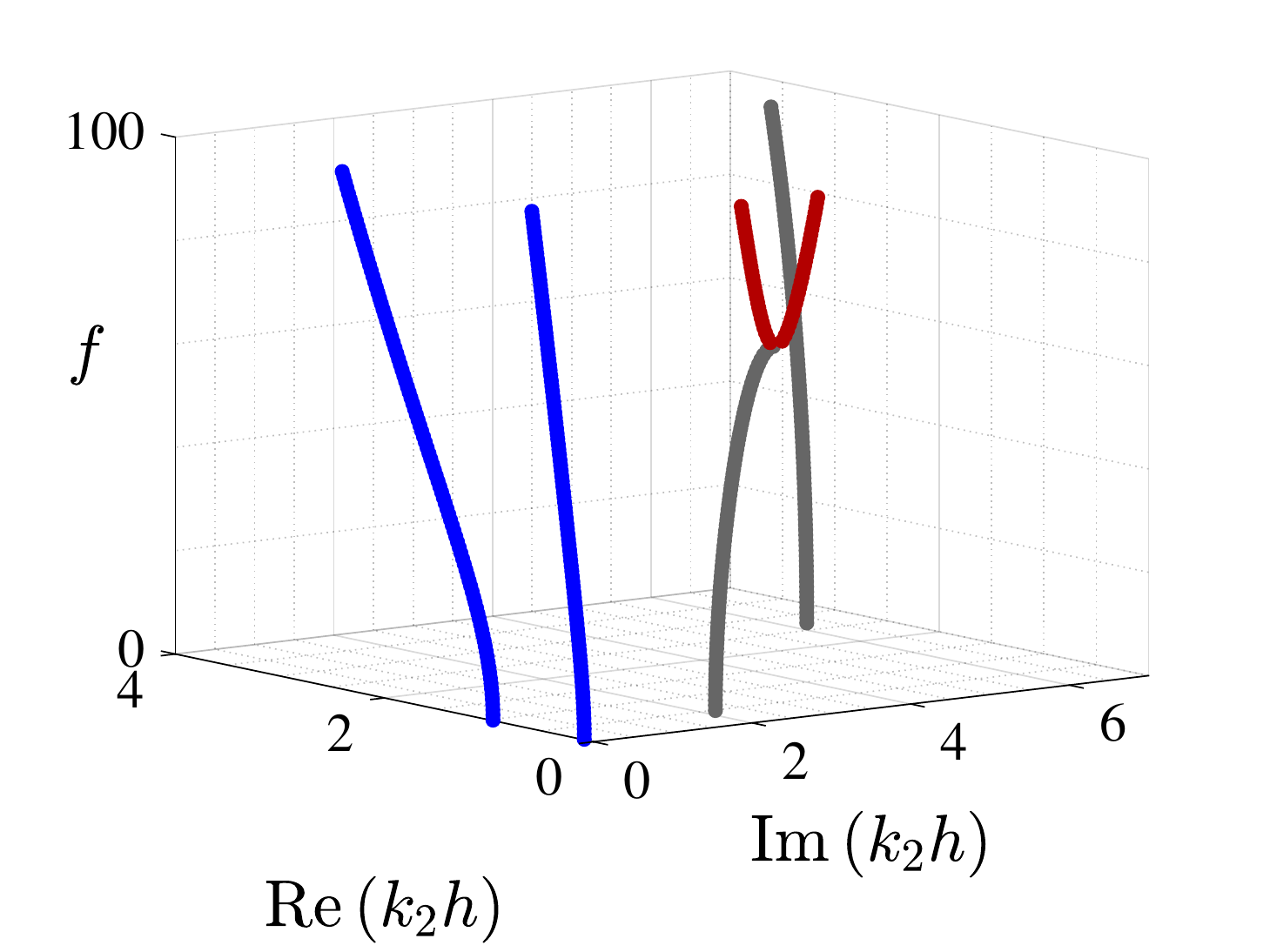}} \quad \sidesubfloat[]{\includegraphics[width=0.25\textwidth]{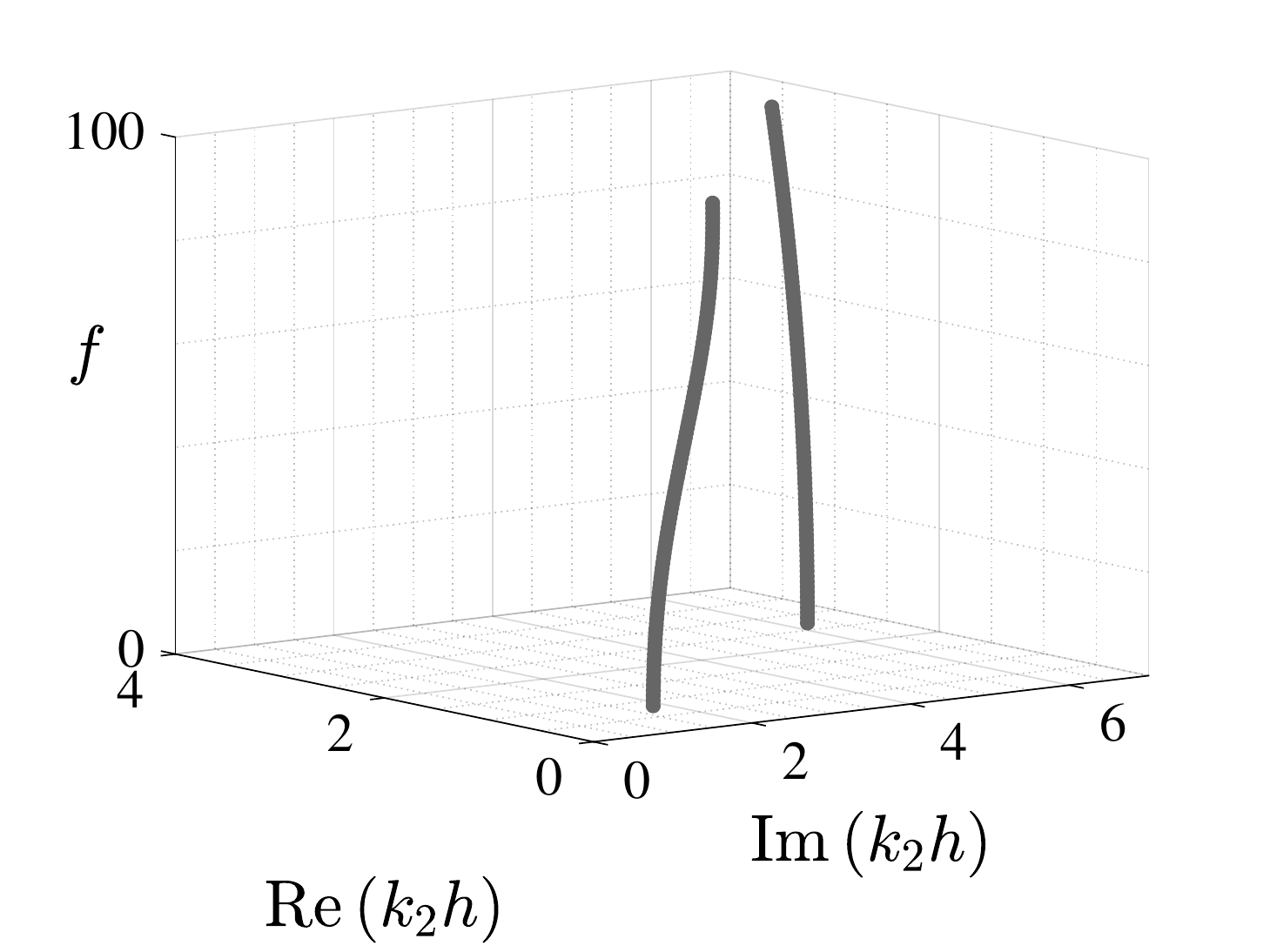}}\caption{The ordinary frequency $f$ (in $\mathrm{kHz}$) versus real and imaginary
parts of $\protect\ksnell h$ solutions for laminate \eqref{eq:lam1}
at (a) $\protect\kb h=0.5$, (b) $0.5i$, and (c) $0.5+0.5i$. Segments
of real, imaginary and complex $\protect\ksnell$ are denoted by blue,
red, and grey, respectively. }

{\small{}{}\label{fig omega vs k2}}{\small\par}
\end{figure}
Finally, in Fig.~\ref{fig omega vs k2} we fix $\kb h$ and evaluate
the spectrum in the $f-\ksnell$ space. Specifically, panels ~\ref{fig omega vs k2}(a)-(c)
correspond to $\kb h=0.5,\,0.5i$ and $0.5+0.5i$, respectively; segments
of real, imaginary and complex $\ksnell$ are denoted by blue, red,
and grey, respectively. We observe that the number of modes in general—and
propagating in particular—increases with frequency. 

In panel \ref{fig omega vs k2}(a) we observe an exceptional point
when $\ksnell h=0.9i$ at 13$\,$kHz, where a single complex mode
branches to two imaginary modes; as frequency increases to 33$\,$kHz,
one of these branches becomes a propagating mode\footnote{\citet{laude2009prb} discussed how the shifting of modes between
the complex and real domains serves as a mechanism to preserve the
total number of modes at a prescribed frequency. }. A similar exceptional point is identified in panel \ref{fig omega vs k2}(b),
when $\ksnell h=2.3i$ at $72\,$kHz. There are no propagating modes
in panel \ref{fig omega vs k2}(c).

\section{Transmission across an interface of a semi-infinite laminate\label{sec:Transmission-across-an}}

As mentioned, modes with either prescribed $\kb$ or prescribed $\ksnell$
are excitable through two different configurations. We analyze the
transmission and excited modes in these configurations, linking them
to the exceptional points, negative refraction, and beam steering
found in the spectrum, as reported in Sec.$\ $\ref{sec:Frequency-Spectrum}. 

\subsection{Metamaterial  refraction via an interface parallel to the layers }

To excite modes with designated vertical lengths, we connect the laminate
at $x_{1}=0$ to a homogeneous half-space whose properties are denoted
by the script 0, from which an incident wave propagates in an angle
$\thetain$ towards the interface (Fig.~\ref{representative lam hom}b).
The wave is described by the potential $\hpoten$
\begin{equation}
\hpoten=\IL e^{i\left(\omega t-\k\sin\thetain x_{1}-\k\cos\thetain x_{2}\right)},\label{eq:potentialhom}
\end{equation}
where the wavenumber $\k=\omega/\hom c$ is either $\omega/\cSz$
for a shear wave, or $\omega/\hom{c_{L}}$ for a pressure wave. The
incident displacements are derived according to Eq.~\eqref{eq:helmholtz},
by setting $\hom{\psi}=\hpoten$ for incident shear, or $\hom{\phi}=\hpoten$
for incident pressure wave. The incident wave is partially reflected
back by the interface, and partially transmitted to the laminate as
Bloch modes. The continuity conditions at $x_{1}=0$ enforces all
these waves to have an $x_{2}$-dependency in the form $e^{-i\k\cos\thetain x_{2}}$.
Corresponding Bloch modes are extracted by the procedure in Sec.~\ref{subsec:Hybrid-Matrix-Method}
with $\ksnell=\k\cos\thetain$, in the form 
\begin{equation}
\T=\sum_{m=1}^{2}\Tm m\upmb{}^{\left(m\right)}\left(x_{1}\right)e^{i\left(\omega t-\kb x_{1}-\k\cos\thetain x_{2}\right)},\label{eq:transmittedparralel}
\end{equation}
where the reflected waves are derived from 
\begin{equation}
\hom{\phiR}=\Rnphi{}e^{i\left(\omega t+\kL 0x_{1}-\k\cos\thetain x_{2}\right)},\ \hom{\psiR}=\Rnpsi{}e^{i\left(\omega t+\kS 0x_{1}-\k\cos\thetain x_{2}\right)},\label{eq:homparralel}
\end{equation}
such that $\kL 0$ and $\kS 0$ are related to $\ksnell=\k\cos\thetain$
via Eq.~\eqref{eq:klksk2}. Finally, the reflection coefficients
$\Rnphi l,\Rnpsi l$ and transmission coefficients $\Tm m$ are determined
from the continuity of the state vector at $x_{1}=0$.

We numerically demonstrate how negative refraction is obtained, using
a half-space with the properties 
\begin{equation}
\nmu 0=0.178\,\mathrm{GPa},\ \nlambda 0=0.714\,\mathrm{GPa},\ \nrho 0=3000\,\mathrm{kg/m^{3}}.\label{eq:hom2-1}
\end{equation}
We set an incident pressure wave at $\thetain=77.22^{\circ}$ and
$f=100\,\mathrm{kHz}$, for which $\ksnell h=1$. The resultant Bloch
wavenumbers are extractable from Fig.~\ref{fig prop modes}, where
at the pertinent $\ksnell$ there is one attenuating mode and one
propagating mode. We recall that the slope in Fig.~\ref{fig prop modes}
of the propagating mode at this $\ksnell$ is positive, while at higher
$\ksnell$ the slope is negative, which is an indicator of negative
refraction. This is verified by evaluating Eqs.~\eqref{eq:ptexplicit}-\eqref{eq:poexplicit},
to find that the propagating angle is $\theta=107.03{}^{\circ}$.
As a consistency check, we examine the balance of energy at $x_{1}=0$,
namely, 
\begin{equation}
\frac{1}{\left|\IL\right|^{2}\sub{\avgP 1}I}\left[\sum_{m=1}^{2}\left|\Tm m\right|^{2}\avgP 1_{T_{m}}^{\left(1\right)}-\left|\Rnphi{}\right|^{2}\avgP 1_{\phi}-\left|\Rnpsi{}\right|^{2}\avgP 1_{\psi}\right]=1,\label{eq: energy conservation parallel layers}
\end{equation}
where the scripts $T_{m},\ \phi,\ \psi$ and $I$ denote quantities
associated with the corresponding fields. Indeed, we find that the
terms in left hand side are $0.617,\ 0.316$ and $0.067$, respectively,
which sum to $1$. Note that only the propagating mode contributes
to the first term, since the energy flux of the mode with the imaginary
$\kb$ vanishes.

As a second example, we set an incident shear incident wave at $\thetain=89.62{}^{\circ}$
and $f=340\,\mathrm{kHz}$, for which $\ksnell h=0.25$. The resultant
Bloch wavenumbers are extractable from Fig.$\ $\ref{fig prop modes}(f).
Here again, at the prescribed $\ksnell$ there is one attenuating
mode, and one propagating mode with a positive slope, that will change
sign at greater $\ksnell$. Calculation of Eqs.~\eqref{eq:ptexplicit}-\eqref{eq:poexplicit}
confirms a negative refraction at $\theta=106.2^{\circ}$. Energy
balance is also verified, where the calculation of the terms in the
left hand side of Eq.~\eqref{eq: energy conservation parallel layers}
provides  $0.148,\ 0.007$ and $0.845$, respectively, which again
sum to $1$.

\begin{figure}[t]
\includegraphics[width=1\textwidth]{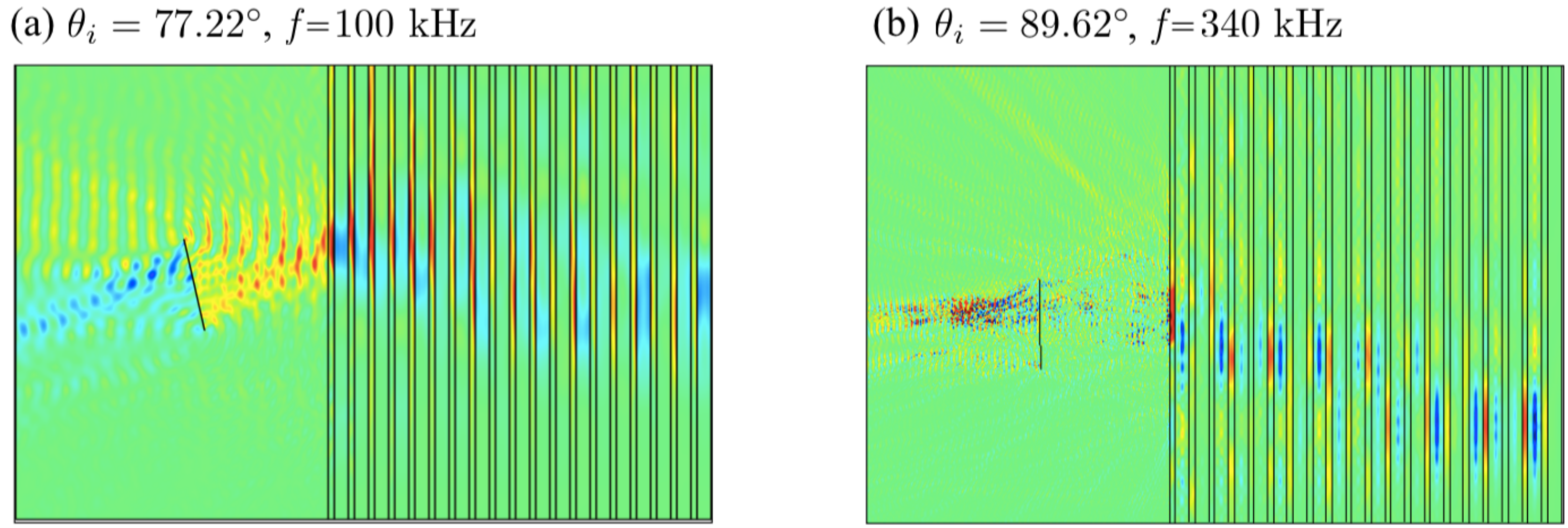}\caption{Frequency domain finite element simulations using COMSOL Multiphysics$^{\circledR}$
of the energy flux along $x_{2}$ ($\mathcal{P}_{2}$) resulting from
an (a) incident pressure wave at $\protect\thetain=77.22^{\circ}$
and $f=100\,\mathrm{kHz}$, and (b) an incident shear incident wave
at $\protect\thetain=89.62{}^{\circ}$ and $f=340\,\mathrm{kHz}$.
The obtained negative angles of refraction agree with the analytical
prediction.}

{\small{}{}\label{fig:FE}}{\small\par}
\end{figure}

Fig.$\ $\ref{fig:FE} shows frequency domain finite element simulations
using COMSOL Multiphysics$^{\circledR}$ the energy flux along $x_{2}$
($\mathcal{P}_{2}$) resulting from an (a) incident pressure wave
at $\thetain=77.22^{\circ}$ and $f=100\,\mathrm{kHz}$, and (b) an
incident shear incident wave at $\thetain=89.62{}^{\circ}$ and $f=340\,\mathrm{kHz}$.
We used low reflecting boundary conditions to avoid reflections due
to the finite truncation of the spatial domain. The incident wave
was excited using load control over a defined line. We used triangular
elements, whose maximal size was third of the unit-cell thickness.
Indeed, the resultant negative angles of refraction agree with the
analytical prediction.

\floatsetup[figure]{style=plain,subcapbesideposition=top}

\subsection{Negative refraction via an interface normal to the layers\label{subsec:tangetboundary}}

It is possible to excite designated Bloch wavenumbers using an incident
wave from a homogeneous half-space that is bonded to the laminate
by an interface along the lamination direction, say at $x_{2}=0$
(Fig.~\ref{representative lam hom}c). This configuration was proposed
by \citet{willis2013arxivB} to achieve negative refraction of anti-plane
shear waves, and was studied in greater detail by \citet{Willis2016jmps,NematNasser2015,Srivastava2016jmps},
and \citet{MORINI2019jmps}. We recall that in this setup, an incident
wave will excite an infinite number of decaying waves in the laminate,
required to enforce continuity conditions \citep{Srivastava2017PRA}.
Here, we firstly develop a method to resolve the normal mode decomposition
and energy partition of the excited in-plane waves in this configuration.

Consider again an incident wave according to the potential \eqref{eq:potentialhom},
now impinging on the horizontal interface $x_{2}=0$. In this setting,
the continuity conditions at the interface enforce the reflected and
transmitted waves to share the same horizontal length with the incident
wave. Accordingly, the transmitted waves have the form 
\begin{equation}
\T=\sum_{m=1}^{\infty}\Tm m\upmb{}^{\left(m\right)}\left(x_{1}\right)e^{i\left(\omega t-\kb x_{1}-\km x_{2}\right)},\label{eq:Tsum}
\end{equation}
where $\kb=k\sin\thetain$, and $\left\{ \km\right\} $ are extractable
from Eq.~\eqref{eq: ev EPWE}. In view of the foregoing observations,
the reflected waves is constructible from potentials that must have
the form
\begin{equation}
\phiR=\sum_{l=-\infty}^{\infty}\Rnphi l\U le^{i\left(\omega t-\kb x_{1}+\knphi lx_{2}\right)},\ \psiR=\sum_{l=-\infty}^{\infty}\Rnpsi l\U le^{i\left(\omega t-\kb x_{1}+\knpsi lx_{2}\right)},\label{potR}
\end{equation}
where
\begin{equation}
\knphi l=\sqrt{\left(\omega/\cLz\right)^{2}-\left(\k\sin\theta+2\pi l/h\right)^{2}},\ \knpsi l=\sqrt{\left(\omega/\cSz\right)^{2}-\left(\k\sin\theta+2\pi l/h\right)^{2}},\label{eq:wavenumberskl}
\end{equation}
and $\U l=e^{-2l\pi ix_{1}/h}$.

The transmission and reflection coefficients $\left\{ \Tm m,\Rnphi l,\Rnpsi l\right\} $
are determined by the continuity conditions at $x_{2}=0$, which are
compactly written as \textbf{
\begin{equation}
\sum_{m=1}^{\infty}\Tm m\m{\stateBC T}-\sum_{l=-\infty}^{\infty}\Rnphi l\l{\stateBC{\phi}}-\sum_{l=-\infty}^{\infty}\Rnpsi l\l{\stateBC{\psi}}=\IL\stateBC I,\label{eq:continouityx2}
\end{equation}
}using 
\begin{equation}
\n{\stateBC{}}=\left(\n{\tilde{u}_{1}},\n{\tilde{u}_{2}},\n{\tilde{\sigma}_{21}},\n{\tilde{\sigma}_{22}}\right)^{\mathsf{T}}.\label{eq:secondstatevector}
\end{equation}
For computational purposes, we truncate the sums over $l$ and $m$
such that $-\NR\leq l\leq\NR$ and $1\leq m\leq2\left(2\NR+1\right)$,
and apply the elegant orthogonality relation\footnote{Alternatively, we can use the Fourier orthogonality $\int_{0}^{h}\U{\alpha}U_{\beta}^{*}\left(x_{1}\right)dx_{1}=h\delta_{\alpha\beta}$.}
\citep{Mokhtari2019arxiv} 
\begin{equation}
\int_{0}^{h}\left[\bb{\stateBC 1}\aa{\conj{\stateBC 3}}+\bb{\stateBC 2}\aa{\conj{\stateBC 4}}-\left(\bb{\stateBC 3}\aa{\conj{\stateBC 1}}+\bb{\stateBC 4}\aa{\conj{\stateBC 2}}\right)\right]\mathrm{d}x_{1}=\frac{4ih}{\omega}\aa{\avgP 2}\delta_{\alpha\beta},\label{eq:orthogonality}
\end{equation}
to obtain an algebraic system of equations for $\left\{ \Tm m,\Rnphi l,\Rnpsi l\right\} $
in the form 
\begin{equation}
\left[\begin{array}{lll}
\M{TT} & \M{T\phi} & \M{T\psi}\\
\M{\phi T} & \M{\phi\phi} & \Zeros{}\\
\M{\psi T} & \Zeros{} & \M{\psi\psi}
\end{array}\right]\left(\begin{array}{l}
\Tm m\\
\Rnphi l\\
\Rnpsi l
\end{array}\right)=\IL\left(\begin{array}{l}
\Ivec T\\
\Ivec{\phi}\\
\Ivec{\psi}
\end{array}\right),\label{eq: system of eq}
\end{equation}
with diagonal submatrices whose elements are given in \nameref{Appendix-M components};
thus, the size of $\M{TT}$ is $2\left(2\NR+1\right)\times2\left(2\NR+1\right)$,
of $\M{T\phi}$ and $\M{T\psi}$ is $2\left(2\NR+1\right)\times\left(2\NR+1\right)$,
of $\M{\phi T}$ and $\M{\psi T}$ is $\left(2\NR+1\right)\times2\left(2\NR+1\right)$,
and the size of the remaining submatrices is $\left(2\NR+1\right)\times\left(2\NR+1\right)$.
In summary, Eq.~\eqref{eq: system of eq} constitutes $4\left(2\NR+1\right)$
algebraic equations for $4\left(2\NR+1\right)$ scattering coefficients.
 \uline{} The resultant coefficients should satisfy the following
energy balance 
\begin{equation}
\frac{1}{\left|\IL\right|^{2}\sub{\avgP 2}I}\left[\sum_{m}\left|\Tm m\right|^{2}\avgP 2_{T}^{\left(m\right)}-\sum_{l}\left|\Rnphi l\right|^{2}\avgP 2_{\phi}^{\left(l\right)}-\sum_{l}\left|\Rnpsi l\right|^{2}\avgP 2_{\psi}^{\left(l\right)}\right]=1.\label{eq: energy conservation}
\end{equation}
An example of the fields at the interface as obtained from the method
is provided in \nameref{Appendix-mode matching}.

\begin{figure}[t]
\centering\sidesubfloat[]{\includegraphics[width=0.35\textwidth]{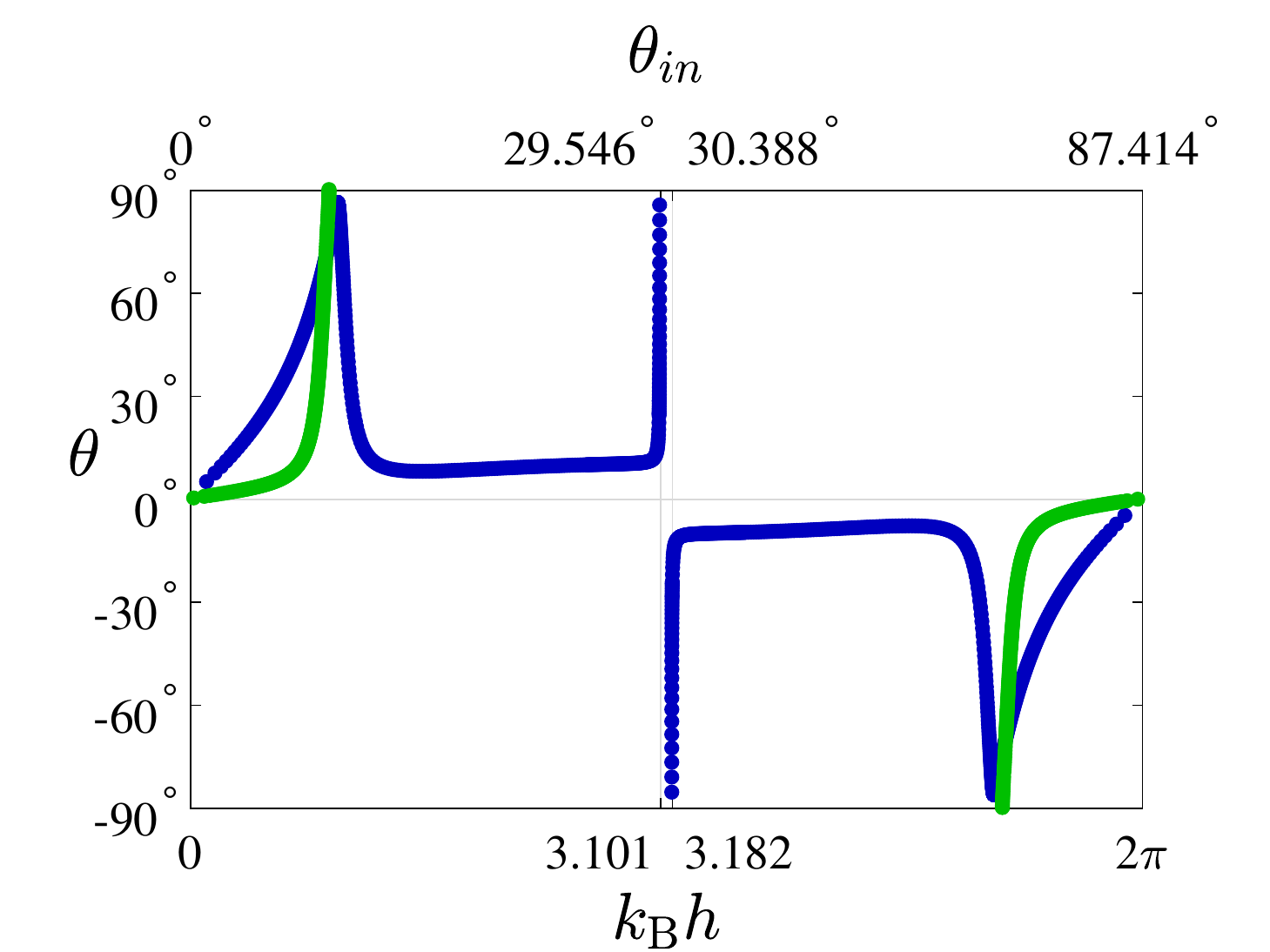}} \quad \sidesubfloat[]{\includegraphics[width=0.35\textwidth]{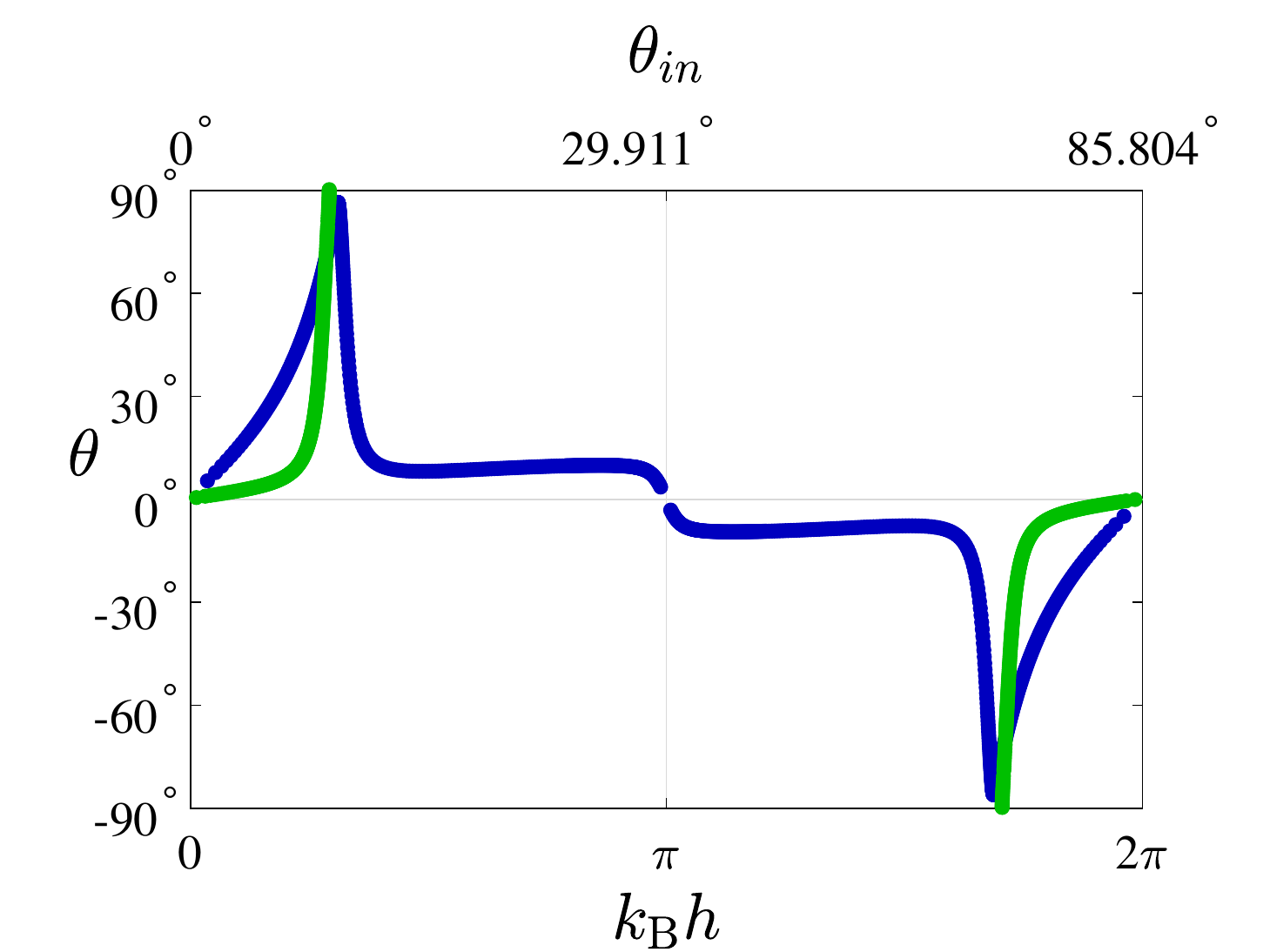}} \quad \caption{The transmitted angle $\protect\thetat$ of Bloch modes in laminate
\eqref{eq:lam1}, induced by a pressure wave impinging at $x_{2}=0$
from the half-space \eqref{eq:hom angles}, as function of incoming
angle $\protect\thetain$ (upper axis) and its corresponding $\protect\kb h$
(lower axis). The frequency is $60\,$kHz in panel (a), and $60.1\,$kHz
in panel (b). The colors identify corresponding branches of the same
color in Fig.~\ref{fig prop modes}. }

{\small{}{}\label{fig:theta vs theta in and kbh}}{\small\par}
\end{figure}
We proceed to analyze the transmission of Bloch modes in laminate
\eqref{eq:lam1} that are induced by impinging pressure wave from
the half-space
\begin{equation}
\nmu 0=0.033\,\mathrm{GPa},\ \nlambda 0=0.132\,\mathrm{GPa},\ \nrho 0=3000\,\mathrm{kg/m^{3}}.\label{eq:hom angles}
\end{equation}
To this end, we plot in Fig.~\ref{fig:theta vs theta in and kbh}
the propagation angle $\theta$ of the modes analyzed in Fig.~\ref{fig prop modes},
as functions of the incoming angle $\thetain$ and its corresponding
$\kb h$, indicated by the upper and lower axes, respectively. The
calculation of $\theta$ was carried out through direct evaluation
of Eqs.~\eqref{eq:ptexplicit}-\eqref{eq:poexplicit} for calculate
$\arctan\frac{\avgP 1}{\avgP 2}$.

Panel \ref{fig:theta vs theta in and kbh}(a) is for $f=60\,$kHz,
which corresponds to Fig.~\ref{fig prop modes}(a), and uses the
same color legend. Our first observation concerns the change of sign
in $\theta$ with respect to $\kb h=\pi$, owing to the $2\pi$-periodicity
of the spectrum in Re$\,\kb h$, and its and reflection symmetry within
that Brillouin zone period. Accordingly, $\partial\omega/\partial\kb$
changes sign, and hence so does $\avgP 1$, as noted by \citet{Willis2016jmps}
in his analysis of anti-plane shear. The diagram exhibits 2-fold rotational
symmetry about $\left(\pi,0\right)$, which corresponds to $\thetain\approx30^{\circ}$.
Notably, the transmission angle of the blue mode increases very fast
near $\kb h=3.101$, and then discontinuously flips from $90^{\circ}$
to $-90^{\circ}$ at $\kb h=3.182$. This discontinuity occurs since
the blue branch does not exist (a gap) when  $3.101<\kb h<\pi$, as
highlighted in Fig. \ref{fig prop modes}(a). Therefore, in the vicinity
$\thetain\approx30^{\circ}$, it is sufficient to slightly change
the incident angle in order to significantly steer the transmitted
mechanical beam. Wide steering by small changes of the incident angle,
albeit less extreme, occurs also about $\thetain\approx10^{\circ}$
and $55^{\circ}$. We further note that since there is a range of
$\kb h$—and hence of incident angles—in which there is only one propagating
mode (see Sec.~\ref{sec:Frequency-Spectrum}), it is possible to
achieve purely negative transmission. Analogous observations were
made by \citet{Srivastava2016jmps} for anti-plane shear waves.

Panel \ref{fig:theta vs theta in and kbh}(b) is for $f=60.1\,$kHz,
which corresponds to Fig.~\ref{fig prop modes}(b). We observe that
the discontinuous sign flip of $\theta$ and significant beam steering
are lost, as the $\theta\left(\kb h\right)$-curve smoothly passes
through the point $\left(0,\pi\right)$. This result agrees with the
extension of the blue branch in Fig.~\ref{fig prop modes}(b) to
$\pi$ near $\ksnell h=0$, where it has an imaginary part, hence
becomes attenuating. Thus, in the vicinity of $f=60\,\mathrm{kHz}$,
energy transport is also very sensitive to the excitation frequency. 

\begin{figure}[t]
\includegraphics[width=0.8\textwidth]{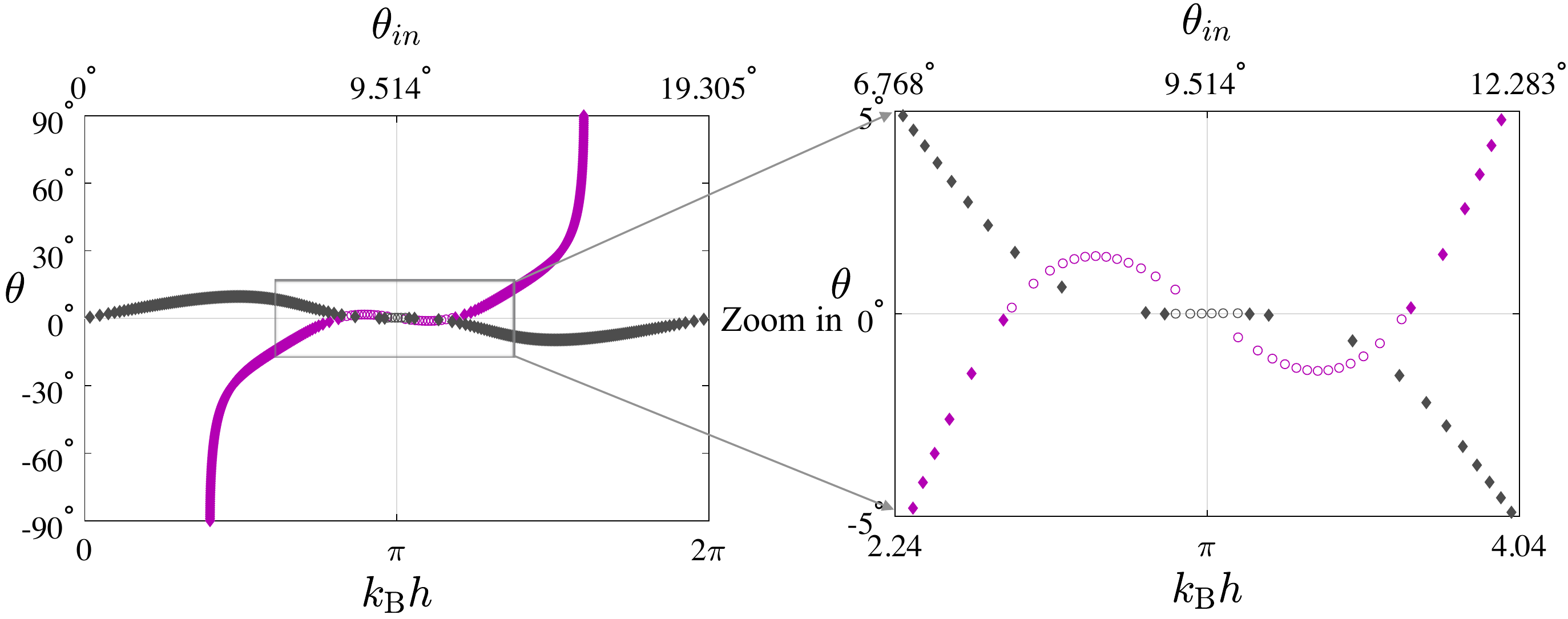}\caption{The transmitted angle $\protect\thetat$ of Bloch modes in laminate
\eqref{eq:lam1}, induced by a pressure wave impinging at $x_{2}=0$
from the half-space \eqref{eq:hom angles}, as function of incoming
angle $\protect\thetain$ (upper axis) and its corresponding $\protect\kb h$
(lower axis), $f=181.3\,$kHz. Purple denotes the solutions with $\protect\ksnell h<0.5$,
where grey denotes solutions with $0.93<\protect\ksnell h<1.21$.
Segments above and below the exceptional points Fig.~\ref{fig prop modes}(e)
are marked using circle marks and diamond marks, respectively. }

{\small{}{}\label{fig:ttheta kB EP}}{\small\par}
\end{figure}
 Fig.~\ref{fig:ttheta kB EP} is for $f=181.3\,$kHz, corresponding
to Fig.~\ref{fig prop modes}(e), which we recall exhibits two exceptional
points; a zoom in about these points is shown in the right side of
the figure. The color legend is different here: purple denotes solutions
with $\ksnell h<0.5$, \emph{i.e.}, left to the first exceptional
point in Fig.~\ref{fig prop modes}(e), where grey denotes solutions
right to the second exceptional point, with $0.93<\ksnell h<1.21$.
(Note that we do not include the third propagating branch with $3.4<\ksnell h$.)
We mark the segments above and below the exceptional points Fig.~\ref{fig prop modes}(e)
using circle marks and diamond marks, respectively. 

Up to the first exceptional point $\left(\kb h\approx2.564,\thetain\approx7.753^{\circ}\right)$,
the green branch of Fig.~\ref{fig prop modes}(e) admits two propagation
angles which—remarkably—one of them is negative \emph{inside} the
first Brillouin zone. Beyond the exceptional point, mode switching
occurs between the green branch (purple diamonds) and the blue branch
(purple circles) of Fig.~\ref{fig prop modes}(e), as the former
becomes attenuating and the latter becomes propagating with positive
refraction. Similar switching occurs between the green and blue branches
beyond the second exceptional point $\left(\kb h\approx3.021,\thetain\approx9.146^{\circ}\right)$.
These phenomena are unique to the in-plane motion, owing to exceptional
points, and are not accessible in the anti-plane setting. Finally,
we note that here again, the diagram exhibits 2-fold rotational symmetry
about $\left(\pi,0\right)$, which corresponds to $\thetain\approx9.514^{\circ}$,
followed by the inversion of the aforementioned phenomena with respect
to the incident angle.s

\section{\label{sec:Conclusions}Conclusions}

We have revisited the problem of in-plane waves propagation in periodic
laminates, were the motivation was twofold. Firstly, this study is
a necessary complement to the reports of \citet{Willis2016jmps,NematNasser2015}
and \citet{Srivastava2016jmps} on metamaterial phenomena in the model
problem of anti-plane waves traversing elastic laminates. Secondly,
this problem allows us to show that the coupling between shear and
pressure parts can be harnessed for anomalous energy transport, within
a relatively simple analytical study. 

We have shown that the corresponding spectrum contains exceptional
points at which two Bloch modes coalesce, and further showed that
these points are the source of anomalous energy transport. To this
end, we have determined how energy is scattered when an incident in-plane
wave impinges the laminate in two model interface problems. We found
metamaterial transmission through the laminate, such as pure negative
refraction, and beam splitting and steering, at states of the system
 near the exceptional points. Notably, we achieved negative refraction
in the canonical transmission problem, where the laminate is impinged
by an incoming wave from a homogeneous medium whose interface with
the laminate is parallel to the layers. 

We emphasize again that these phenomena emerge from the unique coupling
in elastodynamics between the volumetric and distortional modes of
deformation, which cannot be observed in sound and light waves. Our
work further paves the way for encircling exceptional points in a
tangible, purely elastic apparatus, for applications such as asymmetric
mode switches.

\section*{Acknowledgments}

We are grateful to Profs.~Nimrod Moiseyev, Alexei Mailybaev and Ankit
Srivastava for fruitful discussions. We thank Dr.$\ $Pernas-Salomón
for his help regarding the hybrid matrix method. We acknowledge the
support of the Israel Science Foundation, funded by the Israel Academy
of Sciences and Humanities (Grant no.~1912/15), and the United States-Israel
Binational Science Foundation (Grant no.~2014358), and Ministry of
Science and Technology (Grant no.~880011).

\appendix

\global\long\def\theequation{A.\arabic{equation}}%
 \setcounter{equation}{0}

\section*{\label{Appendix HMM}Appendix A}

The forthcoming formulation is an adaptation of the developments of
\citet{Shmuel2016sms} and \citet{pernassalomon2018jmps} to the current
problem. Here, the modified state vector---consisting of quantities
which are continuous across the interface between two adjacent layers---is

\begin{equation}
\sm\left(x_{1},x_{2}\right)=\left(u_{1}\left(x_{1},x_{2}\right)\Cu 1,\ \sigma_{11}\left(x_{1},x_{2}\right)\Csig{11},\ u_{2}\left(x_{1},x_{2}\right)\Cu 2,\ \sigma_{21}\left(x_{1},x_{2}\right)\Csig{21}\right)^{\mathsf{T}},\label{eq:modified state}
\end{equation}
where the normalizing coefficients required to stabilize subsequent
calculations are
\begin{equation}
\begin{array}{ll}
\Cu 1=\left|\ksnell\right|^{-1}, & \Csig{11}=\left(\lambdaAVG\left|\ksnell\right|\left|\kLAVG\right|\right)^{-1},\\
\Cu 2=\left|\ksnell\right|^{-1}, & \Csig{21}=\left(\muAVG\left|\ksnell\right|\left|\kSAVG\right|\right)^{-1},
\end{array}
\end{equation}
and
\begin{equation}
\begin{array}{l}
\kSAVG=\left(\kS{\emph{a}}+\kS{\emph{b}}\right)/2,\\
\kLAVG=\left(\kL{\emph{a}}+\kL{\emph{b}}\right)/2,\\
\muAVG=\left(\nmu{\emph{a}}+\nmu{\emph{b}}\right)/2,\\
\lambdaAVG=\left[\left(\nlambda{\emph{a}}+\nmu{\emph{a}}+\nlambda{\emph{b}}+\nmu{\emph{b}}\right)/2+\muAVG\right]/2.
\end{array}
\end{equation}
The state vector can be written as follows
\begin{equation}
\sm\left(x_{1},x_{2}\right)=\Qmat\left(x_{1},x_{2}\right)\cdot\left(\ALp n,\ \ASp{\emph{n}},\ \ALm{\emph{n}},\ \ASm{\emph{n}}\right)^{\mathsf{T}},
\end{equation}
where
\begin{equation}
\begin{array}{l}
\Qmat\left(x_{1},x_{2}\right)=\left(1/\Cu 1,\ 1/\Csig{11},\ 1/\Cu 2,\ 1/\Csig{21}\right)\cdot\\
\left(\begin{array}{llll}
i\kL{\emph{p}} & -i\ksnell & -i\kL{\emph{p}} & -i\ksnell\\
-\nlambda{\emph{p}}\ksnell^{2}-\left(\nlambda{\emph{p}}+2\nmu{\emph{p}}\right)\kLSquare{\emph{p}} & 2\nmu{\emph{p}}\kS{\emph{p}}\ksnell & -\nlambda{\emph{p}}\ksnell^{2}-\left(\nlambda{\emph{p}}+2\nmu{\emph{p}}\right)\kLSquare{\emph{p}} & -2\nmu{\emph{p}}\kS{\emph{p}}\ksnell\\
-i\ksnell & -i\kS{\emph{p}} & -i\ksnell & i\kS{\emph{p}}\\
2\nmu{\emph{p}}\kL{\emph{p}}\ksnell & \nmu{\emph{p}}\left(\kSSquare{\emph{p}}-\ksnell^{2}\right) & -2\nmu{\emph{p}}\kL{\emph{p}}\ksnell & \nmu{\emph{p}}\left(\kSSquare{\emph{p}}-\ksnell^{2}\right)
\end{array}\right)\\
\cdot\left(e^{i\kL{\emph{p}}x1},\ e^{i\kS{\emph{p}}x1},\ e^{-i\kL{\emph{p}}x1},\ e^{-i\kS{\emph{p}}x1}\right)e^{-i\ksnell x_{2}}.
\end{array}
\end{equation}
$\Qmat\left(x_{1},x_{2}\right)$ can also be defined as
\begin{eqnarray}
\Qmat\left(x_{1},x_{2}\right) & = & \left(\begin{array}{llll}
\fmat_{1}\left(x_{1},x_{2}\right) & \fmat_{2}\left(x_{1},x_{2}\right) & \fmat_{3}\left(x_{1},x_{2}\right) & \fmat_{4}\left(x_{1},x_{2}\right)\\
\amat_{1}\left(x_{1},x_{2}\right) & \amat_{2}\left(x_{1},x_{2}\right) & \amat_{3}\left(x_{1},x_{2}\right) & \amat_{4}\left(x_{1},x_{2}\right)
\end{array}\right).
\end{eqnarray}
Using this definition the hybrid matrix of the $\emph{n}^{\mathrm{th}}$
layer is given by 
\begin{equation}
\begin{aligned}\nH{\emph{n}}\left(\nh{\emph{p}}\right)= & \left(\begin{array}{llll}
\fmat_{1}\left(x_{0},x_{2}\right) & \fmat_{2}\left(x_{0},x_{2}\right) & \fmat_{3}\left(x_{0},x_{2}\right) & \fmat_{4}\left(x_{0},x_{2}\right)\\
\amat_{1}\left(x_{0}+\nh{\emph{p}},x_{2}\right) & \amat_{2}\left(x_{0}+\nh{\emph{p}},x_{2}\right) & \amat_{3}\left(x_{0}+\nh{\emph{p}},x_{2}\right) & \amat_{4}\left(x_{0}+\nh{\emph{p}},x_{2}\right)
\end{array}\right)\\
 & \left(\begin{array}{llll}
\amat_{1}\left(x_{0},x_{2}\right) & \amat_{2}\left(x_{0},x_{2}\right) & \amat_{3}\left(x_{0},x_{2}\right) & \amat_{4}\left(x_{0},x_{2}\right)\\
\fmat_{1}\left(x_{0}+\nh{\emph{p}},x_{2}\right) & \fmat_{2}\left(x_{0}+\nh{\emph{p}},x_{2}\right) & \fmat_{3}\left(x_{0}+\nh{\emph{p}},x_{2}\right) & \fmat_{4}\left(x_{0}+\nh{\emph{p}},x_{2}\right)
\end{array}\right)^{-1}.
\end{aligned}
\label{eq: H of single layer}
\end{equation}
Using the definition in Eq.$\,$\eqref{eq: H of single layer}, we
relate the field variables which appear in the modified state vector
at the ends $x_{0}$ and $x_{0}+\nthc{\emph{p}}$ of the $\emph{n}^{\mathrm{th}}$
layer via 
\begin{equation}
\left(\begin{array}{l}
\n{u_{1}}\left(x_{0},x_{2}\right)\Cu 1\\
\n{\sigma_{11}}\left(x_{0},x_{2}\right)\Csig{11}\\
\n{u_{2}}\left(x_{0}+\nh{\emph{p}},x_{2}\right)\Cu 2\\
\n{\sigma_{21}}\left(x_{0}+\nh{\emph{p}},x_{2}\right)\Csig{21}
\end{array}\right)=\nH{\emph{n}}\left(\nh{\emph{p}}\right)\cdot\left(\begin{array}{l}
\n{u_{2}}\left(x_{0},x_{2}\right)\Cu 2\\
\n{\sigma_{21}}\left(x_{0},x_{2}\right)\Csig{21}\\
\n{u_{1}}\left(x_{0}+\nh{\emph{p}},x_{2}\right)\Cu 1\\
\n{\sigma_{11}}\left(x_{0}+\nh{\emph{p}},x_{2}\right)\Csig{11}
\end{array}\right),
\end{equation}
where $\nH{\emph{n}}\left(\nh{\emph{p}}\right)$ is the $4\times4$
hybrid matrix of the $\emph{n}^{\mathrm{th}}$ layer. In order to
calculate the total hybrid matrix of $\emph{n}$ layers, denoted by
$\nH{1,n}$, a recursive algorithm is used in terms of the hybrid
matrix of the first $\emph{n}-1$ layers ($\nH{1,n-1}$) and the hybrid
matrix of the $n^{\mathrm{th}}$ layer $\nH{\emph{n}}$, namely, \begin{equation}
\begin{alignedat}{1}\bH_{22}^{(1,k)} & =\bH_{22}^{\left(k\right)}+\bH_{21}^{\left(k\right)}\cdot\left[\bI-\bH_{22}^{(1,k-1)}\cdot\bH_{11}^{\left(k\right)}\right]^{-1}\cdot\bH_{22}^{(1,k-1)}\cdot\bH_{12}^{\left(k\right)},\\
\bH_{21}^{(1,k)} & =\bH_{21}^{\left(k\right)}\cdot\left[\bI-\bH_{22}^{(1,k-1)}\cdot\bH_{11}^{\left(k\right)}\right]^{-1}\cdot\bH_{21}^{\left(1,k-1\right)},\\
\bH_{12}^{(1,k)} & =\bH_{12}^{(1,k-1)}\cdot\bH_{12}^{\left(k\right)}+\bH_{12}^{(1,k-1)}\cdot\bH_{11}^{\left(k\right)}\cdot\left[\bI-\bH_{22}^{(1,k-1)}\cdot\bH_{11}^{\left(k\right)}\right]^{-1}\cdot\bH_{22}^{(1,k-1)}\cdot\bH_{12}^{\left(k\right)},\\
\bH_{11}^{(1,k)} & =\bH_{11}^{(1,k-1)}+\bH_{12}^{(1,k-1)}\cdot\bH_{11}^{\left(k\right)}\cdot\left[\bI-\bH_{22}^{(1,k-1)}\cdot\bH_{11}^{\left(k\right)}\right]^{-1}\cdot\bH_{21}^{(1,k-1)}, 
\end{alignedat} 
\label{eq: composition rule for H} 
\end{equation}Where $\bH_{11}^{\left(1,k\right)},\bH_{12}^{\left(1,k\right)},\bH_{21}^{\left(1,k\right)},\bH_{22}^{\left(1,k\right)}$
denote the $2\times2$ blocks of the hybrid matrix $\bH^{\left(1,k\right)}$
of the first $k$ layers. The generalized eigenproblem presented in
Eq. \eqref{eq:ev problem} yields the characteristic equation 
\begin{equation}
a_{1}+\lambda\a 2+\lambda^{2}\a 3+\lambda^{3}\a 4+\lambda^{4}\a 5=0\label{eq:polinom ev}
\end{equation}
for $\lambda=e^{i\kb h}$, where \begin{equation} 	
\begin{aligned} 
&a_{1}=h_{13} h_{24}-h_{14} h_{23}=\mathrm{det}\,
\mathsf{H}_{12},\\	
&a_{2}=-h_{13} h_{42} h_{31}-h_{24} h_{42} h_{31}+h_{12} h_{43} h_{31}+h_{22} h_{44} h_{31}-h_{31}+h_{13} h_{32} h_{41}\\
&+h_{24} h_{32} h_{41}-h_{12} h_{33} h_{41}-h_{22} h_{34} h_{41}\\ 	
&+h_{11} h_{33} h_{42}+h_{21} h_{34} h_{42}-h_{42}-h_{11} h_{32} h_{43}-h_{21} h_{32}  h_{44},\\
&a_{3}=1+h_{13} h_{31}+h_{24} h_{31}-h_{14} h_{23} h_{42} h_{31}+h_{13} h_{24} h_{42} h_{31}+h_{14} h_{22} h_{43} h_{31}\\ 	
&-h_{12} h_{24} h_{43} h_{31}-h_{13} h_{22} h_{44} h_{31}+h_{12} h_{23} h_{44} h_{31}-h_{11} h_{33}-h_{21} h_{34}\\ 
&+h_{14} h_{23} h_{32} h_{41}-h_{13} h_{24} h_{32} h_{41}-h_{14} h_{22} h_{33} h_{41}+h_{12} h_{24} h_{33} h_{41}+h_{13} h_{22}h_{34} h_{41}\\ 	
&-h_{12} h_{23} h_{34} h_{41}+h_{13} h_{42}+h_{24} h_{42}+h_{14} h_{21} h_{33} h_{42}-h_{11} h_{24} h_{33} h_{42}\\ 
&-h_{13} h_{21} h_{34} h_{42}+h_{11} h_{23} h_{34} h_{42}-h_{12} h_{43}-h_{14} h_{21} h_{32} h_{43}+h_{11} h_{24} h_{32} h_{43}\\ 
&+h_{12} h_{21} h_{34} h_{43}-h_{11} h_{22} h_{34} h_{43}-h_{22} h_{44}+h_{13} h_{21} h_{32} h_{44}-h_{11} h_{23} h_{32} h_{44}\\ 	
&-h_{12} h_{21} h_{33} h_{44}+h_{11} h_{22} h_{33} h_{44},\\
&\a 5=\a 1,\ \a 4=\a 2,	
\end{aligned} 
\end{equation}and $h_{ij}$ are the components of the total hybrid matrix. The latter
two equalities imply that $\lambda^{-1}=e^{-i\kb h}$ is also a solution,
as expected. Eq.~\eqref{eq:polinom ev} yields a closed-form expression
for $\cos\kb h$, given in Eq. \eqref{eq: coskbh}. 

\global\long\def\theequation{B.\arabic{equation}}%
 \setcounter{equation}{0}

\section*{\label{Appendix- EPWE}Appendix B}

The matrices \textbf{$\Amat_{0},\Amat_{1},\Amat_{2}$ }and $\Bmat$
appearing in Eq. \eqref{eq: ev PWE} are given by

\begin{equation}
\begin{aligned}\Amat_{0}=\left(\begin{array}{ll}
\Amat_{0}^{11} & \Zeros{}\\
\Zeros{} & \Amat_{0}^{22}
\end{array}\right),\ \Amat_{1}=\left(\begin{array}{ll}
\Zeros{} & \Amat_{1}^{12}\\
\Amat_{1}^{21} & \Zeros{}
\end{array}\right),\ \Amat_{2}=\left(\begin{array}{ll}
\Amat_{2}^{11} & \Zeros{}\\
\Zeros{} & \Amat_{2}^{22}
\end{array}\right),\ \Bmat=\left(\begin{array}{ll}
\Bmat^{11} & \Zeros{}\\
\Zeros{} & \Bmat^{22}
\end{array}\right)\end{aligned}
,
\end{equation}
where the $\left(\G,\GTag\right)$ component of each block is 
\begin{equation}
\begin{aligned} & \Amat_{0\ \G\GTag}^{11}=-\left[\left(\G+\kb\right)\left(\GTag+\kb\right)\right]\left[\lambda\left(\G-\GTag\right)+2\mu\left(\G-\GTag\right)\right],\\
 & \Amat_{0\ \G\GTag}^{22}=-\left[\left(\G+\kb\right)\left(\GTag+\kb\right)\right]\mu\left(\G-\GTag\right),\\
 & \Amat_{1\ \G\GTag}^{12}=\Amat_{1\ \G,\GTag}^{21}=-\left(\G+\kb\right)\left[\lambda\left(\G-\GTag\right)+\mu\left(\G-\GTag\right)\right],\\
 & \Amat_{2\ \G\GTag}^{11}=-\mu\left(\G-\GTag\right),\\
 & \Amat_{2\ \G\GTag}^{22}=-\lambda\left(\G-\GTag\right)-2\mu\left(\G-\GTag\right),\\
 & \Bmat_{\G\GTag}^{11}=\Bmat_{\G,\GTag}^{22}=-\rho\left(\G-\GTag\right).
\end{aligned}
\end{equation}


\global\long\def\theequation{C.\arabic{equation}}%
 \setcounter{equation}{0}

\section*{\label{Appendix-COMPARISON EPWE AND HMM}Appendix C}

\begin{figure}[t]
\centering\includegraphics[width=0.7\textwidth]{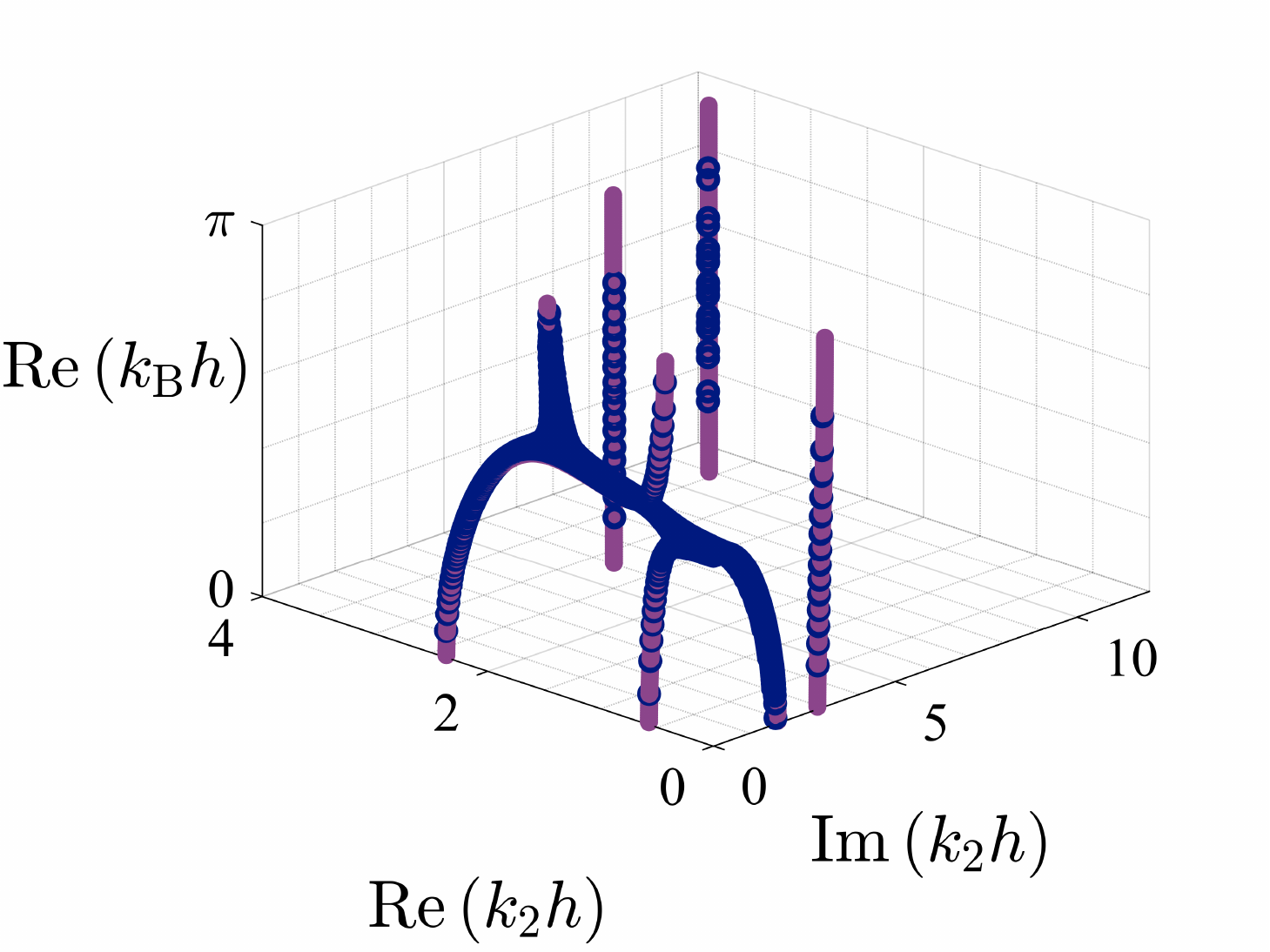}\caption{real $\protect\kb h$ as function of the real and imaginary parts
of $\protect\ksnell h$ for laminate \eqref{eq:lam1} and $f=100\ \mathrm{kHz}$,
using the EPWE and Hybrid matrix method, depicted in purple and blue
points, respectively.}

{\small{}{}\label{fig hmm vs epwe}}{\small\par}
\end{figure}
This appendix demonstrate the applicability of the extended plane
wave expansion method, by comparing its results with the exact results
of the hybrid matrix method. Fig. \ref{fig hmm vs epwe} shows real
$\kb h$ as function of the real and imaginary parts of $\ksnell h$
for laminate \eqref{eq:lam1} and $f=100\ \mathrm{kHz}$. Points calculated
with the extended plane wave expansion method and hybrid matrix method
are depicted in purple and blue, respectively. We used 51 plane waves
for the extended plane wave expansion method and points calculated
using the hybrid matrix method differ from the results of the hybrid
matrix by $5\times10^{-2}$.

\global\long\def\theequation{B.\arabic{equation}}%
 \setcounter{equation}{0}

\section*{\label{Appendix-M components}Appendix D}

The components of the submatrices in Eq.~\ref{eq: system of eq}
are

\begin{equation}
\begin{aligned} & \M{TT_{mm}}=\frac{4ih}{\omega}\avgP 2_{T}^{\left(m\right)},\\
 & \M{T\phi_{ml}}=\int_{0}^{h}\left[\l{\stateBC{\phi_{1}}}\m{\conj{\stateBC{T_{3}}}}+\l{\stateBC{\phi_{2}}}\m{\conj{\stateBC{T_{4}}}}-\left(\l{\stateBC{\phi_{3}}}\m{\conj{\stateBC{T_{1}}}}+\l{\stateBC{\phi_{4}}}\m{\conj{\stateBC{T_{2}}}}\right)\right]\mathrm{d}x_{1},\\
 & \M{T\psi_{ml}}=\int_{0}^{h}\left[\l{\stateBC{\psi_{1}}}\m{\conj{\stateBC{T_{3}}}}+\l{\stateBC{\psi_{2}}}\m{\conj{\stateBC{T_{4}}}}-\left(\l{\stateBC{\psi_{3}}}\m{\conj{\stateBC{T_{1}}}}+\l{\stateBC{\psi_{4}}}\m{\conj{\stateBC{T_{2}}}}\right)\right]\mathrm{d}x_{1},\\
 & \M{\phi T_{lm}}=\int_{0}^{h}\left[\m{\stateBC{T_{1}}}\l{\conj{\stateBC{\phi_{3}}}}+\m{\stateBC{T_{2}}}\l{\conj{\stateBC{\phi_{4}}}}-\left(\m{\stateBC{T_{3}}}\l{\conj{\stateBC{\phi_{1}}}}+\m{\stateBC{T_{4}}}\l{\conj{\stateBC{\phi_{2}}}}\right)\right]\mathrm{d}x_{1},\\
 & \M{\phi\phi_{ll}}=\frac{4ih}{\omega}\avgP 2_{\phi}^{\left(l\right)},\\
 & \M{\psi T_{lm}}=\int_{0}^{h}\left[\m{\stateBC{T_{1}}}\l{\conj{\stateBC{\psi_{3}}}}+\m{\stateBC{T_{2}}}\l{\conj{\stateBC{\psi_{4}}}}-\left(\m{\stateBC{T_{3}}}\l{\conj{\stateBC{\psi_{1}}}}+\m{\stateBC{T_{4}}}\l{\conj{\stateBC{\psi_{2}}}}\right)\right]\mathrm{d}x_{1},\\
 & \M{\psi\psi_{ll}}=\frac{4ih}{\omega}\avgP 2_{\psi}^{\left(l\right)},\\
 & \Ivec{T_{m}}=\int_{0}^{h}\left[\stateBC{I_{1}}\m{\conj{\stateBC{T_{3}}}}+\stateBC{I_{2}}\m{\conj{\stateBC{T_{4}}}}-\left(\stateBC{I_{3}}\m{\conj{\stateBC{T_{1}}}}+\stateBC{I_{4}}\m{\conj{\stateBC{T_{2}}}}\right)\right]\mathrm{d}x_{1},\\
 & \Ivec{\phi_{l}}=\int_{0}^{h}\left[\stateBC{I_{1}}\l{\conj{\stateBC{\phi_{3}}}}+\stateBC{I_{2}}\l{\conj{\stateBC{\phi_{4}}}}-\left(\stateBC{I_{3}}\l{\conj{\stateBC{\phi_{1}}}}+\stateBC{I_{4}}\l{\conj{\stateBC{\phi_{2}}}}\right)\right]\mathrm{d}x_{1},\\
 & \Ivec{\psi_{l}}=\int_{0}^{h}\left[\stateBC{I_{1}}\l{\conj{\stateBC{\psi_{3}}}}+\stateBC{I_{2}}\l{\conj{\stateBC{\psi_{4}}}}-\left(\stateBC{I_{3}}\l{\conj{\stateBC{\psi_{1}}}}+\stateBC{I_{4}}\l{\conj{\stateBC{\psi_{2}}}}\right)\right]\mathrm{d}x_{1},
\end{aligned}
\end{equation}
where subscript numbers denote the component of $\stateBC{}$, and
subscript letters denote the potential from which this  component
is derived. 


\global\long\def\theequation{B.\arabic{equation}}%
 \setcounter{equation}{0}

\section*{\label{Appendix-mode matching}Appendix E}

\floatsetup[figure]{style=plain,subcapbesideposition=top}

\begin{figure}[t]
\centering\sidesubfloat[]{\includegraphics[width=0.35\textwidth]{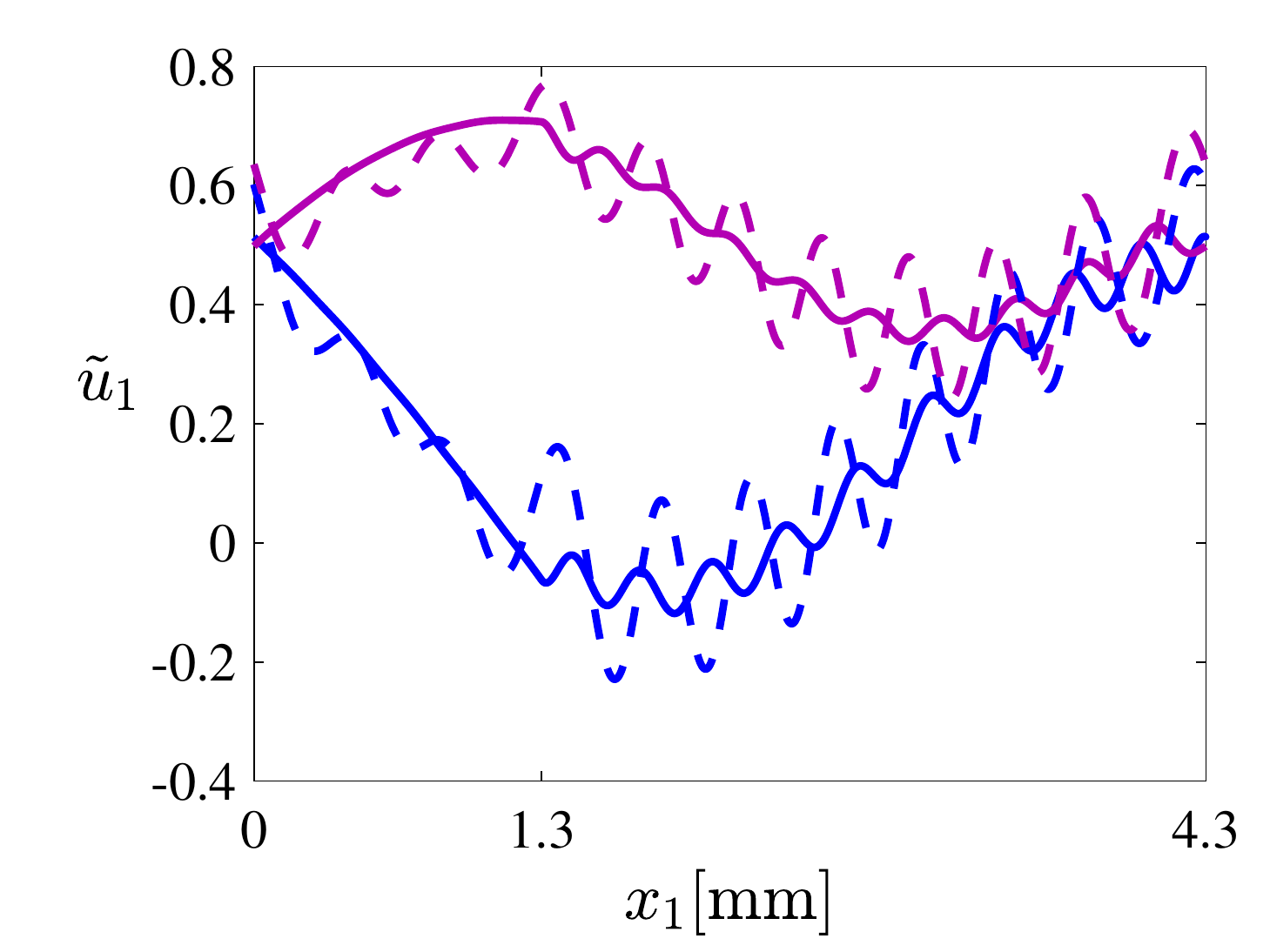}} \quad \sidesubfloat[]{\includegraphics[width=0.35\textwidth]{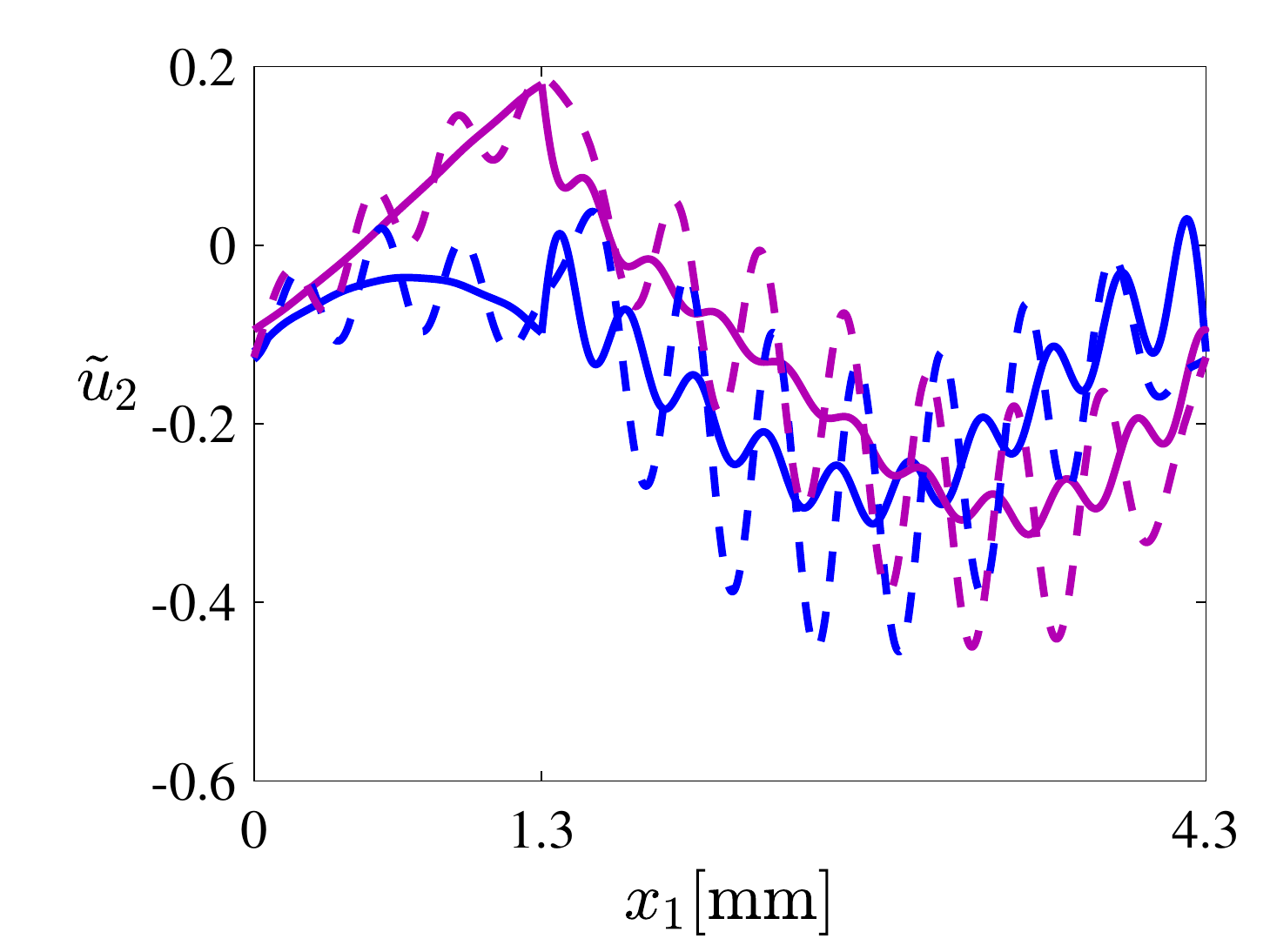}} \quad \sidesubfloat[]{\includegraphics[width=0.35\textwidth]{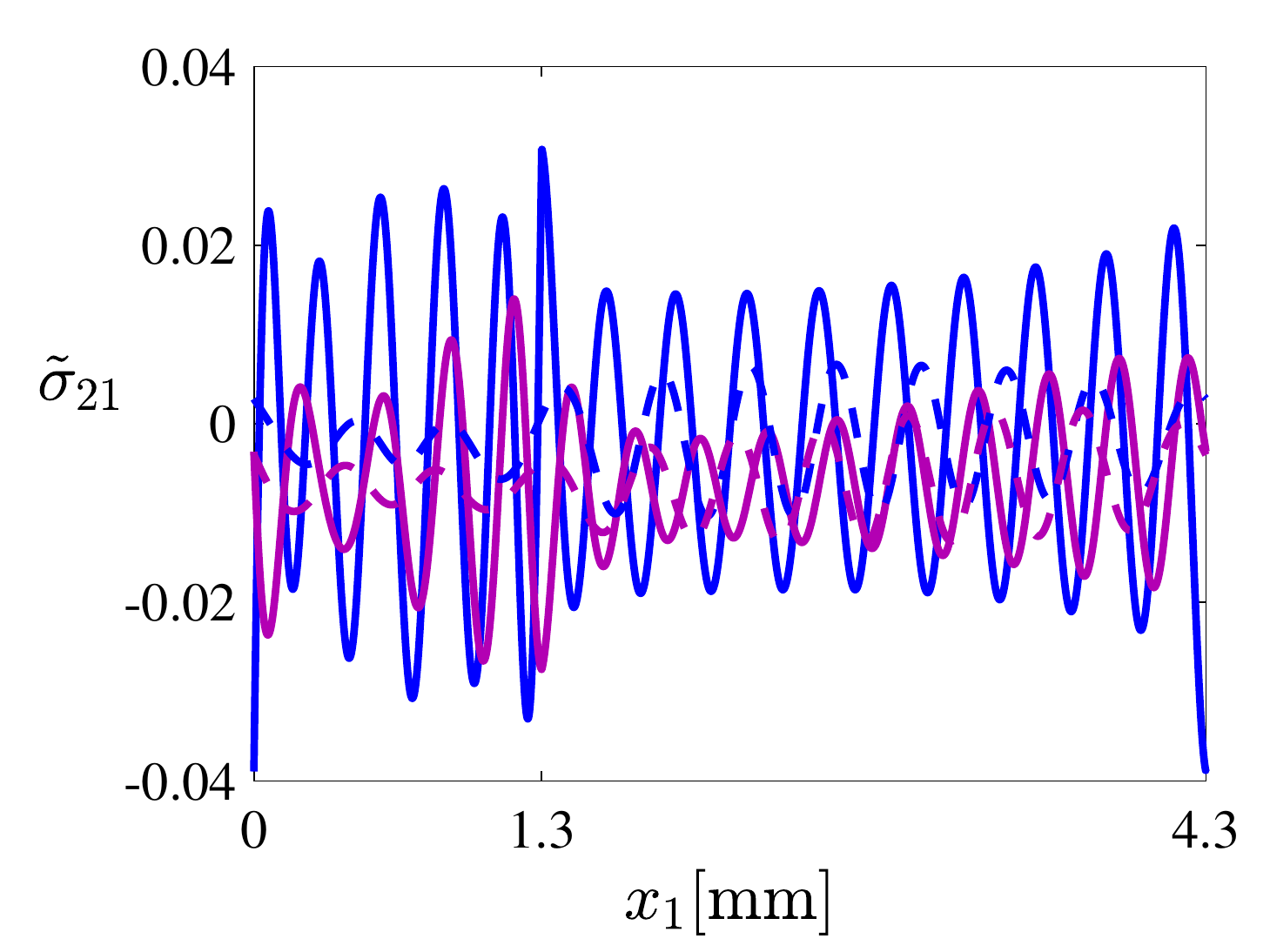}} \quad \sidesubfloat[]{\includegraphics[width=0.35\textwidth]{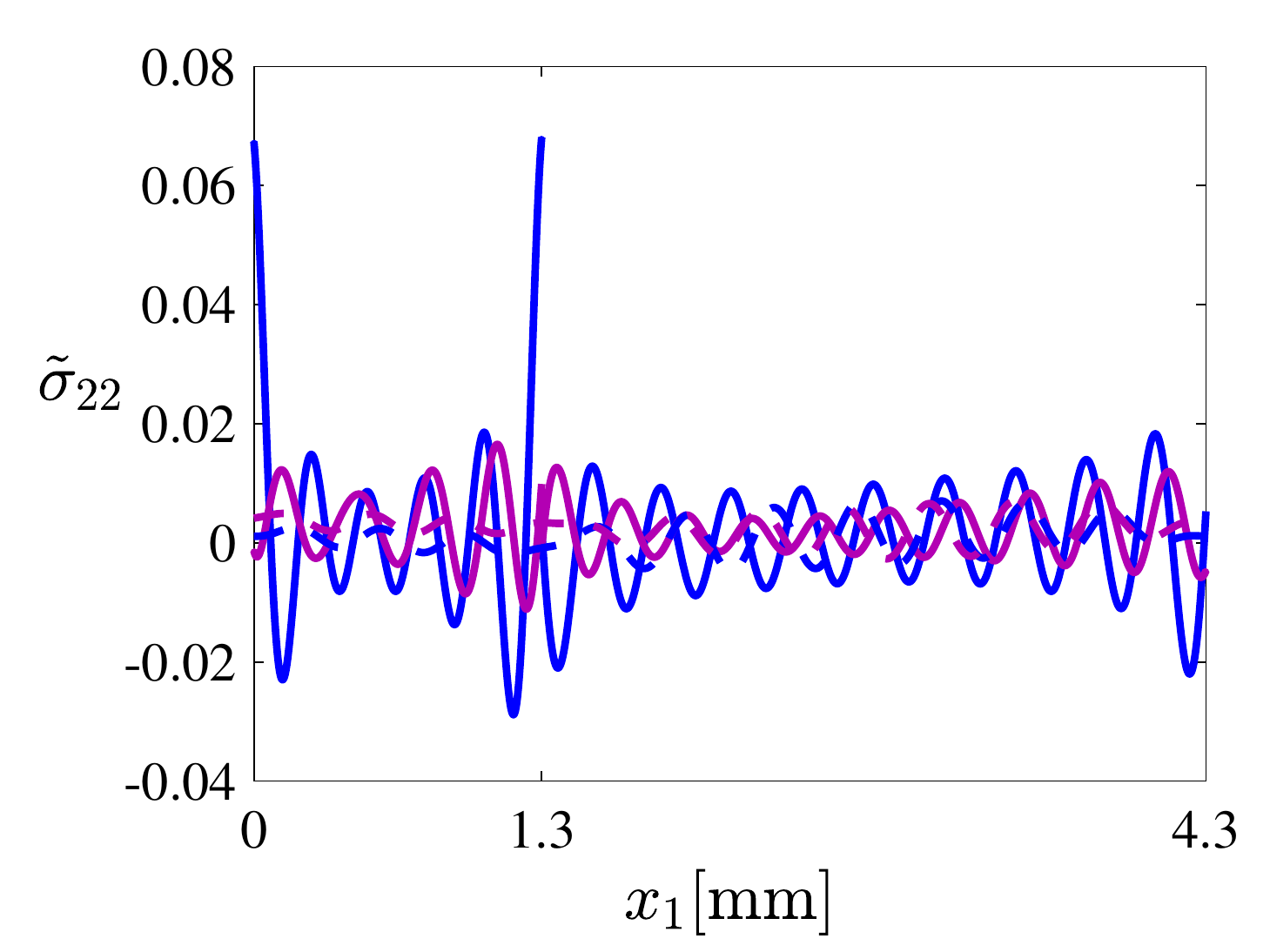}}\caption{(a) $\protect\stateBC 1$ (b) $\protect\stateBC 2$ (c) $\protect\stateBC 3$
(d) $\protect\stateBC 4$ at the interface $x_{2}=0$ between laminate
\eqref{eq:lam1} and the homogeneous half-space \eqref{eq:hom2-1},
for $f=100\ \mathrm{kHz}$, $\protect\thetain=15.72^{\circ}$ an incident
shear wave. The fields in the laminate (resp.$\ $homogeneous half-space)
fields are depicted in continuous (dashed) curves, where the real
(imaginary) part is depicted in blue (purple).}

{\small{}{}\label{fig lam 1 - scattering}}{\small\par}
\end{figure}
Fig. \ref{fig lam 1 - scattering} shows $\stateBC 1$ (panel a),
$\stateBC 2$ (panel b), $\stateBC 3$ (panel c, normalized by $k\muAVG$),
and $\stateBC 4$ (panel d, normalized by $k\lambdaAVG$) at the interface
$x_{2}=0$ as calculated via Eq.$\,$\eqref{eq: system of eq}. Specifically,
solid and dashed curves correspond to the laminate and homogeneous
half-space, respectively, where the real and imaginary parts of each
field are denoted by blue and purple, respectively. The laminate and
homogeneous half-space that was used have properties \eqref{eq:lam1}
and \eqref{eq:hom2-1}, respectively, and the calculation was carried
out for $f=100\,\mathrm{kHz}$, $\thetain=15.72^{\circ}$, and an
incident shear wave. For this example, we chose $\NR=12$, which ensures
all the propagating transmitted and reflected modes in the $x_{2}$
direction are included; the rest of the transmitted modes are those
to have the lowest absolute value of $\mathrm{Im}\ksnell h$. The
solution obtained does not match the fields perfectly, however it
is sufficient on average provides a good approximation. The solution
yields the following terms  for the energy balance \eqref{eq: energy conservation},
\begin{equation}
\begin{array}{ll}
\left|\Tm 1\right|^{2}\avgP 2_{T}^{\left(1\right)}\slash\left|\IL\right|^{2}\sub{\avgP 2}I=0.049, & \left|\Rnpsi{-1}\right|^{2}\avgP 2_{\psi}^{\left(-1\right)}\slash\left|\IL\right|^{2}\sub{\avgP 2}I=-0.0133,\\
\left|\Rnphi{-1}\right|^{2}\text{\ensuremath{\avgP 2_{\phi}^{\left(-1\right)}}}\slash\left|\IL\right|^{2}\sub{\avgP 2}I=-0.091, & \left|\Rnpsi 0\right|^{2}\avgP 2_{\psi}^{\left(0\right)}\slash\left|\IL\right|^{2}\sub{\avgP 2}I=-0.437,\\
\left|\Rnphi 0\right|^{2}\avgP 2_{\phi}^{\left(0\right)}\slash\left|\IL\right|^{2}\sub{\avgP 2}I=-0.391, & \left|\Rnpsi 1\right|^{2}\avgP 2_{\psi}^{\left(1\right)}\slash\left|\IL\right|^{2}\sub{\avgP 2}I=-0.007,\\
\left|\Rnpsi{-2}\right|^{2}\avgP 2_{\psi}^{\left(-2\right)}\slash\left|\IL\right|^{2}\sub{\avgP 2}I=0,
\end{array}
\end{equation}
which sum to  0.988: a difference of only 0.12\%. In this case,
the transmitted mode refracts positively with transmitted angle of
$\theta=1.12^{\circ}$, however carries only a small fraction of the
energy of the incident wave. 

\bibliographystyle{plainnat}

\end{document}